\documentclass[showpacs,amsmath,amssymb,twocolumn,
nofootinbib]{revtex4}

\begin{document}
\title{Selection rules for the Wheeler-DeWitt equation in quantum cosmology}
\author{A.O. Barvinsky}
\affiliation{Theory Department, Lebedev
Physics Institute, Leninsky Prospect 53, Moscow 119991, Russia}
\author{A.Yu. Kamenshchik}
\affiliation{Dipartimento di Fisica e Astronomia and INFN, Via Irnerio 46, 40126 Bologna,
Italy\\
L.D. Landau Institute for Theoretical Physics of the Russian
Academy of Sciences,\\ Kosygin Str. 2, Moscow 119334, Russia}

\begin{abstract}
Incompleteness of Dirac quantization scheme leads to a redundant set of solutions of the Wheeler-DeWitt equation for the wavefunction in superspace of quantum cosmology. Selection of physically meaningful solutions matching with quantum initial data can be attained by a reduction of the theory to the sector of true physical degrees of freedom and their canonical quantization. The resulting physical wavefunction unitarily evolving in the time variable introduced within this reduction can then be raised to the level of the cosmological wavefunction in superspace of 3-metrics to form a needed subset of all solutions of the Wheeler-DeWitt equation. We apply this technique in several simple but nonlinear minisuperspace models and discuss both at classical and quantum level physical reduction in {\em extrinsic} time -- the time variable determined in terms of extrinsic curvature (or momentum conjugated to the cosmological scale factor). Only this extrinsic time gauge can be consis!
 tently used in vicinity of turning points and bounces where the scale factor reaches extremum and cannot monotonically parameterize the evolution of the system. Since the 3-metric scale factor is canonically dual to extrinsic time variable, the transition from the physical wavefunction to the wavefunction in superspace represents a kind of the generalized Fourier transform. This transformation selects square integrable solutions of the Wheeler-DeWitt equation, which guarantee Hermiticity of canonical operators of the Dirac quantization scheme. This makes this scheme consistent -- the property which is not guaranteed on general solutions of the Wheeler-DeWitt equation. Semiclassically this means that wavefunctions are represented by oscillating waves in classically allowed domains of superspace and exponentially fall off in classically forbidden (underbarrier) regions. This is explicitly demonstrated in flat FRW model with a scalar field having a constant negative potential 
 and for the case of phantom scalar field with a positive potential. The FRW model of a scalar field with a vanishing potential does not lead to selection rules for solutions of the Wheeler-DeWitt equation, but this does not violate Hermiticity properties, because all these solutions are anyway of plane wave type and describe cosmological dynamics without turning points and bounces. In models with turning points the description of classically forbidden domains goes beyond original principles of unitary quantum reduction to the physical sector, because it includes complexification of the physical time variable or complex nature of the physical Hamiltonian. However, this does not alter the formalism of the Wheeler-DeWitt equation which continues describing underbarrier quantum dynamics in terms of {\em real} superspace variables.
\end{abstract}
\pacs{98.80.Qc, 98.80.Jk, 04.60.Ds}
\maketitle

\section{Introduction}
Needless to say that quantum cosmology is an inalienable tool in the studies of the very early quantum Universe. Inflationary cosmology \cite{inflation}, whose observational status was essentially strengthened after the first Planck release \cite{Planck}, successfully resolves many traditional problems of the Big Bang theory, but it is unable to resolve the issue of initial conditions for the cosmological evolution. This issue is an ultimate goal of quantum cosmology.

On the other hand, it would be not a great exaggeration to say that quantum cosmology is one of the most discredited areas in modern theoretical physics. The Wheeler-DeWitt equation \cite{DeWitt}, as an incarnation of this theory, is widely considered as a decorative tool that would never lead to original achievements in gravity theory. At best, it would be used as a justification of the semiclassical results obtained by other much simpler and down-to-earth methods of quantum field theory in curved spacetime. Though much of this criticism seems to be really true, the situation with this equation very much resembles the status of modern $S$-matrix theory vs Schr\"odinger equation and its stationary perturbation theory. Despite the fact that scattering amplitudes are much simpler to obtain by the covariant Feynman diagrammatic technique, everyone clearly understands that their machinery is based on the fundamental but manifestly noncovariant Schr\"odinger equation, and without
  it this technique becomes a set of ungrounded contrived rules.

A similar situation holds for the Wheeler-DeWitt equation which underlies the physical dynamically independent content of the theory and its numerous applications. However, there is a big difference -- the formalism of the Wheeler-DeWitt equation itself, without additional assumptions, does not form a closed physical theory. These assumptions concern two main ingredients of the theory -- selection of physically meaningful solutions of the Wheeler-DeWitt equation and the construction of the physical inner product on their subspace, that could generate observable amplitudes and expectation values.

Matter of fact is that this equation has many more solutions than those having clear physical setup. This setup is dictated by dynamically independent degrees of freedom, which are intricately hidden among the full set of gravitational variables. Various ways to disentangle these degrees of freedom and separate them from the gauge variables give rise to different quantization schemes. One scheme consists in disentangling them at the classical level and canonically quantizing in the resulting reduced phase space. Another scheme is the Dirac quantization \cite{Dirac}, when the first class Dirac constraints are imposed as equations on the quantum states in the representation space in which all original phase space variables (both physical and gauge ones) satisfy canonical commutation relations. This scheme is not closed as a physical theory, because it does not have by itself a well-defined conserved positive definite inner product that would provide unitarity of the theory. Ho!
 wever, exactly this approach is usually employed in numerous applications of quantum cosmology. As a rule they do not go beyond achieving the solution of Wheeler-DeWitt equation and giving it some interpretation or fundamentally restricting the gravitational sector to the semiclassical domain \cite{Rubakov} and using Born-Oppenheimer approximation \cite{BO}.

The most striking feature of this approach is a mismatch between the nature of the Wheeler-DeWitt equation and the number of its boundary (or initial) value data. As a second order differential equation (of the Klein-Gordon type in minisuperspace applications) it requires two data at the Cauchy surface -- the value of the function and its normal (or time) derivative, but any local unitary quantum theory assumes the existence of only one initial data function -- wavefunction of the initial state. This clearly demonstrates incompleteness of the Dirac quantization, and the goal of our work is to achieve its completion by formulating the selection rules for solutions of the Wheeler-DeWitt equation and demonstration of these rules in several simple minisuperspace models.

In the minisuperspace context the redundancy in solutions of this equation directly manifests itself in positive and negative frequency solutions, existing due to the hyperbolic rather than parabolic nature of the Wheeler-DeWitt equation. Usual interpretation of these solutions as describing expanding and contracting cosmological models breaks down in the vicinity of the bounce -- the point of maximal or minimal size of the Universe, where expansion goes over into contraction or vice versa. Consistent description of this situation is possible if we start with the reduced phase space quantization with an appropriately chosen time variable. This automatically leads to the missing selection rule in the set of solutions of the Wheeler-DeWitt equation, and as a by-product raises the issue of Hermiticity of canonically commuting operators in the definition of quantum Dirac constraint.

The structure of the paper is as follows. In Sect. II we present unitarity approach to the Wheeler-DeWitt formalism \cite{BarvU} which allows one to raise the reduced phase space quantum state to the level of the wavefunction in the DeWitt superspace of quantum cosmology. This procedure, described in Sect.III, implies a special {\em time-nonlocal} transform from the physical wavefunction, satisfying a conventional Schr\"odinger equation, to the Wheeler-DeWitt wavefunction and leads to the selection rules of the above type. In the following sections we apply this transform in several minisuperspace models and discuss relevant operator Hermiticity and unitarity properties. In Sects. IV-V we consider the model with a negative constant potential. In Sect. VI we consider the model of a massless scalar field with a vanishing potential. Section VII is devoted to the model, based on the phantom scalar field with a positive constant potential. Section VIII is devoted to discussion an!
 d conclusions. Appendices A and B present useful formulae for unitary canonical transformations and integral representations for Bessel and modified Bessel functions.

\section{Unitarity approach to quantum cosmology}
Gravity theory in the canonical formalism has the action which in condensed DeWitt notations
\cite{DeWitt1} is of the following form
        \begin{eqnarray}
        S=\int dt\,\big\{p_i\dot{q}^i
        -N^\mu H_\mu(q,p)\big\}.                    \label{1.1}
        \end{eqnarray}
Here $q^i$ represent spatial metric coefficients and matter fields. Their conjugated momenta are denoted by $p_i$. The condensed index $i$ includes discrete labels and also {\em spatial} coordinates, and Einstein summation rule implies integration over the latter. The same concerns the indices $\mu$ of the Lagrange multipliers $N^\mu$ which are the ADM lapse and shift functions \cite{ADM}. In the formal treatment of the infinite dimensional configuration space as a finite-dimensional manifold (what we assume in this section) the range of $i$ is $i=1,...n$ and the range of $\mu$ is $\mu=1,...m$ with $n>m$. In asymptotically-flat models the integrand of (\ref{1.1}) contains also the contribution of the ADM surface term Hamiltonian $-H_0(q,p)$, but below we consider only spatially closed or spatially flat minisuperspace models without this term.

The variation of the Lagrange multipliers leads to
the set of nondynamical equations -- constraints
        \begin{eqnarray}
        H_\mu(q,p)=0,                            \label{1.2}
        \end{eqnarray}
which in gravity theory comprise the set of Hamiltonian and momentum constraints. The constraint functions $H_\mu(q,p)$ belong to the first class and satisfy the Poisson-bracket algebra
        \begin{eqnarray}
        &&\{H_\mu,H_\nu\}=U^\lambda_{\mu\nu}H_\lambda,   \label{1.3}
        \end{eqnarray}
with the structure functions $U^\lambda_{\mu\nu}=U^\lambda_{\mu\nu}(q,p)$ which depend on phase-space variables of the theory.

Dirac quantization of the theory (\ref{1.1}) consists in promoting the initial phase-space variables and the constraint functions to the operator level $(q,p,H_\mu)\rightarrow (\hat{q},\hat{p},\hat{H}_\mu)$ and
in selecting the physical states $|\,{\mbox{\boldmath$\varPsi$}}\big>$ in the
representation space of $(\hat{q},\hat{p},\hat{H}_\mu)$ by the equation
        \begin{eqnarray}
        \hat{H}_\mu|\,{\mbox{\boldmath$\varPsi$}}\big>=0.   \label{1.5}
        \end{eqnarray}
The operators $(\hat{q},\hat{p})$ satisfy the canonical commutation relations $[\hat{q}^k,\hat{p}_l]=i\hbar\delta^k_l$ and the quantum constraints $\hat{H}_\mu$ as operator functions of $(\hat{q},\hat{p})$ should satisfy the correspondence principle with the classical $c$-number constraints and be subject to the commutator algebra
        \begin{eqnarray}
        [\hat{H}_\mu,\hat{H}_\nu]=
        i\hbar\hat{U}^{\lambda}_{\mu\nu}
        \hat{H}_\lambda.                          \label{1.6}
        \end{eqnarray}
with some operator structure functions $\hat{U}^{\lambda}_{\mu\nu}$ standing to the left of operator constraints. This algebra generalizes (\ref{1.3}) to the quantum level and serves as integrability conditions for equations (\ref{1.5}).

\subsection{Classical reduction to the physical sector: coordinate gauge conditions}

The theory with the action (\ref{1.1}) is invariant under the set of transformations of phase space variables $(q^i,p_i)$ and Lagrange multipliers \cite{BarvU} signifying diffeomorphism invariance of the original action in the Lagrangian form. Conventional approach to this situation implies that the equivalence class of variables belonging to the orbit of these transformations corresponds to one and the same physical state. The description of this state in terms of physical variables consists in singling out the unique representative of each such class and in treating the
independent labels of this representative as physical variables.

This can be attained by imposing on original phase space variables the gauge conditions
        \begin{eqnarray}
        \chi^\mu(q,p,t)=0                      \label{5.03}
        \end{eqnarray}
which determine in the $2n$-dimensional phase space the $(2n-m)$-dimensional surface (remember that $n$ is the range of index $i$, while $m$ is that of $\mu$) having a unique intersection with the orbit of gauge transformations. At least locally, the latter condition means the invertibility of the Faddeev-Popov matrix \cite{FaddeevPopov} with the nonvanishing determinant
        \begin{eqnarray}
        J=\det J^\mu_\nu,\quad J^\mu_\nu=
        \{\chi^\mu,H_\nu\}.                    \label{5.2}
        \end{eqnarray}
Gauge conditions of the form (\ref{5.03}) impose restrictions only on phase space variables and single out the physical sector locally in time -- all variables in the canonical action
expressed in terms of true physical degrees of freedom. This goes as follows.

To begin with, the Lagrange multipliers (which are not fixed by equations of motion for the action (\ref{1.1})) become uniquely fixed as functions of $(q^i,p_i)$. This directly follows from the conservation in time of gauge conditions, serving as the equation for lapse and shift functions
        \begin{eqnarray}
        &&\frac{d}{dt}\chi^\mu=\{\chi^\mu,H_\nu\}N^\nu
        +\frac{\partial\chi^\mu}{\partial t}=0,\\
        &&N^\mu=-J^{-1\,\mu}_{\,\,\,\;\;\;\,\nu}
        \frac{\partial\chi^\nu}{\partial t}.    \label{lapse0}
        \end{eqnarray}
For the reparameterization invariant  systems with $H_0(q,p)=0$ in (\ref{1.1}), the gauge conditions should explicitly depend on time in order to generate dynamics in the physical phase space \cite{BarvU,BKr}. For gravitational systems with the Lagrangian multipliers playing the role of lapse and shift functions this is obvious -- nonzero values of the latter exist only for $\partial\chi^\nu/\partial t\neq 0$.
\footnote{The geometrical meaning of $N^\mu$ is the collection of normal and tangential projections of 4-velocity with which a spacelike hypersurface moves in the embedding spacetime, so that degeneration of $N^\mu$ to zero implies freezing this surface at a fixed position. Then it does not scan the spacetime and no physical dynamics is probed within this time-independent gauge conditions \cite{BarvU}.}

The parametrization of $(q^i,p_i)$ in terms of phase space variables of the physical sector $(\xi^A,\pi_A)$, $A=1,...n-m$, in its turn, follows from solving together the system of constraints (\ref{1.2}) and gauge conditions (\ref{5.03}), which determine the embedding of the $2(n-m)$-dimensional physical phase space into the space of $(q^i,p_i)$
        \begin{eqnarray}
        &&q^i=q^i(\xi^A,\pi_A,t),\\
        &&p_i=p_i(\xi^A,\pi_A,t).
        \end{eqnarray}
Internal coordinates of this embedding should satisfy the canonical transformation law for the symplectic form restricted to the physical subspace
        \begin{eqnarray}
        p_i dq^i=
        \pi_A d\xi^A
        -H_{\rm phys}(\xi^A,\pi_A,t)dt
        +dF(q^i,\xi^A,t),             \label{sympform}
        \end{eqnarray}
so that $\xi^A$ and $\pi_A$ can be respectively identified with physical phase space coordinates and conjugated momenta, $H_{\rm phys}(\xi^A,\pi_A,t)$ considered as a physical Hamiltonian and $F(q^i,\xi^A,t)$ -- the generating function of this canonical transformation.

The simplest form of this reduction is for gauge conditions imposed only on phase space coordinates, $\chi^\mu(q,t)=0$. Such {\em coordinate} gauge conditions determine the embedding of the $(n-m)$-dimensional space $\Sigma$ of physical {\em coordinates} $\xi^A$ directly into the space of original coordinates $q^i$,
        \begin{eqnarray}
        \Sigma: q^i=e^i(\xi^A,t),\,\,\,
        \chi^\mu(e^i(\xi,t),t)\equiv 0.   \label{5.4}
        \end{eqnarray}
Here $\xi^A$ are identified with the physical coordinates, and the conjugated momenta $\pi_A$ and the physical Hamiltonian reads
        \begin{eqnarray}
        &&\pi_A=p_i\frac{\partial e^i}{\partial\xi^A},\\
        &&H_{\rm phys}(\xi^A,\pi_A,t)=-p_i(\xi^A,\pi_A,t)\frac{\partial e^i(\xi,t)}{\partial t},    \label{5.8a}
        \end{eqnarray}
As we see, the physical momenta are the projections of the original momenta $p_i$ to the vectors tangential to $\Sigma$, $e^i_A\equiv\partial e^i/\partial\xi^A$. In view of the contact nature of this transformation the generating function $F(q^i,\xi^A,t)$ in (\ref{sympform}) is vanishing.

The normal projections of $p_i$ should be found from the
constraints (\ref{1.2}), the local uniqueness of their solution being granted by the nondegeneracy of the Faddeev-Popov determinant. Together with (\ref{5.4}) this solution yields all the original phase space variables $(q^i,p_i)$ as known functions of the physical degrees of freedom $(\xi^A,\pi_A)$. The original action (\ref{1.1}) reduced to the physical sector (that is to the subspace of constraints and gauge conditions) acquires the usual canonical form with the physical Hamiltonian (\ref{5.8a}).

\subsection{Quantum reduction}
Canonical quantization in the physical sector (or reduced phase space quantization) consists in promoting $\xi^A,\pi_A,H_{\rm phys}(\xi^A,\pi_A,t)$ to the level of operators $\hat\xi^A,\hat\pi_A,\hat H_{\rm phys}$, subject to canonical commutation relations $[\hat\xi^A,\hat\pi_A]=i\hbar\delta^A_B$, and postulating the Schr\"odinger equation for the quantum state of the system in the Hilbert space of these operators
        \begin{eqnarray}
        i\hbar\frac\partial{\partial t}\varPsi_{\rm phys}(t,\xi)=\hat H_{\rm phys}
        \varPsi_{\rm phys}(t,\xi).    \label{SchroedEq}
        \end{eqnarray}
Here $\varPsi_{\rm phys}(t,\xi)=\langle\xi|\varPsi_{\rm phys}(t)\rangle$ is the wave function of this state in the coordinate representation.
The kernel of its unitary evolution can be represented by the path integral over trajectories in the reduced phase space,
\begin{eqnarray}
        &&\varPsi_{\rm phys}(t_+,\xi_+)
        =\int d\xi_-\nonumber\\
        &&\qquad\qquad\times
        K(t_+,\xi_+|t_-,\xi_-)\,
        \varPsi_{\rm phys}(t_-,\xi_-),      \label{5.10a}\\
        \nonumber\\
&&K(t_+,\xi_+|t_-,\xi_-)
=\int\limits_{\xi(t_\pm)=\xi_\pm}D[\xi,\pi]\nonumber\\
&&\qquad\quad\times\exp\frac{i}{\hbar}
        \int\limits_{t_-}^{t_+}dt
        \big\{\pi_A\dot\xi^A
        -H_{\rm phys}(\xi,\pi,t)\big\},   \label{unitaryK}
        \end{eqnarray}
where $D[\xi,\pi]=\prod_t d\xi(t)\,d\pi(t)$ is the Liouville measure of integration over trajectories interpolating between the two points $\xi_\pm$ -- the arguments of the evolution kernel.

Profound success in quantization of gauge theories achieved in seventies and eighties of the last century \cite{FaddeevPopov,DeWitt,DeWitt1,FV,BFV} was based on the identical transformation in this path integral from the variables of the reduced phase space to the variables of the original action (\ref{1.1}). This transformation brings us to the Faddeev-Popov canonical path integral for the two-point kernel in the space of original coordinates $q^i$ -- DeWitt superspace of 3-matrics and matter fields \cite{BarvU}
        \begin{eqnarray}
        &&\!\!\!{\mbox{\boldmath$K$}}(q_+,q_-)
        =\int\limits_{q(t_\pm)=q_\pm}
        D[\,q,p\,]\,DN\,\nonumber\\
        &&\!\!\!\times\prod\limits_{t_+>t>t_-} J_t\,\delta(\chi_t)\,\exp\frac{i}{\hbar}
        \int\limits_{t_-}^{t_+}dt
        \big\{p_i\dot q^i-N^\mu H_\mu\big\},   \label{K}
        \end{eqnarray}
where $D[q,p]=\prod_t \prod_i dq^i(t)\,dp_i(t)$ is the Liouville measure of integration over trajectories interpolating between the points $q_\pm$,
            \begin{eqnarray}
            DN=\prod\limits_{t_+\geq t\geq t_-}
            \prod_{\mu}\, dN^{\mu}(t)
            \end{eqnarray}
is the integration measure over lapse and shift functions including the integration over $N^{\mu}(t_\pm)$ at boundary points $t_\pm$ and $\prod_t J_t\,\delta(\chi_t)$ is the Faddeev-Popov gauge fixing factor
    \begin{equation}
    J_t\,\delta(\chi_t)\equiv \det J^\mu_\nu\big(q(t),p(t)\big)
    \prod\limits_\alpha\delta\big(\chi^\alpha(q(t),t)\big),
    \end{equation}
which restricts the integration over $q(t)$ at any $t\neq t_\pm$ to the gauge condition surface (\ref{5.03}).

Integration over $N^\mu(t_\pm)$ has a very important consequence. It implies that

\begin{equation}
\hat H_\mu\Big(q_+,\frac{\hbar}i\frac{\partial}{\partial q_+}\Big)
{\mbox{\boldmath$K$}}(q_+,q_-) = 0,
\label{WdW-lapse}
\end{equation}
i.e. one finds that the two-point kernel ${\mbox{\boldmath$K$}}$ is a solution of the quantum Dirac constraint $\hat H_\mu {\mbox{\boldmath$K$}}=0$ -- the Wheeler-DeWitt equation.
One can show that due to the well known gauge independence properties of the Faddeev-Popov path integral this kernel is independent of the choice of $\chi^\mu$ (in the class of admissible gauges). Also in view of the time parametrization invariance it is independent of $t_\pm$.\footnote{The kernel (\ref{K}) is a truncation of the BFV unitary evolution operator in the relativistic phase space to the zero ghost sector \cite{BFV,BarvU}, and these properties directly follow from this truncation \cite{WhyBFV}.}

The role of $\mbox{\boldmath$K$}(q,q')$ is revealed by the observation that in the semiclassical approximation it can be related to the unitary evolution operator (\ref{unitaryK})
\cite{GenSem,BKr,BarvU}. This relation derived in \cite{BarvU} by slicing the path integral and confirmed in the semiclassical approximation in \cite{GenSem,BKr} reads
        \begin{eqnarray}
        &&K(t,\xi|t',\xi')\nonumber \\
        &&\quad=
        \left.\left(\frac{\overrightarrow{\!\!J}}M\right)^{1/2}\!\!
        {\mbox{\boldmath$K$}}(q,q')
        \left(\frac{\overleftarrow{J'}}{M'}\right)^{1/2}
        \right|_{\,q=e(\xi,t),\,
        q'=e(\xi',t')}\nonumber \\
        &&\quad+O(\hbar),          \label{5.9}
        \end{eqnarray}
where the operator-valued factors $\overrightarrow{J}$ and $\overleftarrow{J'}$ coincide with the Faddeev-Popov determinants in which the c-number momentum argument is replaced by its operator representation\footnote{Operator ordering in (\ref{operatorJ}) is immaterial, because it is responsible for terms of higher order in $\hbar$ which go beyond the semiclassical approximation including the (one-loop) prefactor.},
        \begin{eqnarray}
        \overrightarrow{\!\!J}=J\Big(q,\frac{\hbar}i
        \frac{\overrightarrow\partial}{\partial q}\Big),\quad
        \overleftarrow{J'}=J\Big(q,-\frac{\hbar}i
        \frac{\overleftarrow
        \partial}{\partial q'}\Big),   \label{operatorJ}
        \end{eqnarray}
and $M=M(q)$ is a measure of integration over the $(n-m)$-dimensional physical subspace $\Sigma$ in $n$-dimensional $q$-space, satisfying   
        \begin{eqnarray}
        d^{n-m}\xi=d^n q\,\delta(\chi(q)) M(q). \label{M}
        \end{eqnarray}

This implies that the kernel ${\mbox{\boldmath$K$}}(q,q')$ similarly to the Schr\"odinger propagator $K(t,\xi|t',\xi')$ can be regarded as a propagator of the Dirac wavefunction ${\mbox{\boldmath$\varPsi$}}(q)$ in
$q$-space. The boundary value problem for this wavefunction can be written down as
        \begin{eqnarray}
        &&\hat H_\mu\Big(q,\frac{\hbar}i\frac{\partial}{\partial q}\Big)\,
        \mbox{\boldmath$\varPsi$}(q)=0,        \label{5.11}\\
        &&\varPsi_{\rm phys}(\xi,t)=\left.
        \left(\frac{\overrightarrow{\!\!J}}
        M\right)^{1/2}\,
        \mbox{\boldmath$\varPsi$}(q)\,
        \,\right|_{\,q=e(\xi,t)}
        \!\!\!\!+O(\hbar),                      \label{5.11a}
        \end{eqnarray}
where the relation (\ref{5.11a}) serves as ``initial" condition for ${\mbox{\boldmath$\varPsi$}}(q)$ specified on the $(n-m)$-dimensional surface $\Sigma$, and $m$-equations (\ref{5.11}), $\mu=1,...m$, propagate this initial data onto entire $n$-dimensional superspace.\footnote{Note that $\Sigma$ is not the hypersurface and its codimension is $m>1$, so that $m$ equations (\ref{5.11}) to recover $\mbox{\boldmath$\varPsi$}(q)$ on the full $n$-dimensional superspace from the boundary data (\ref{5.11a}).} This propagation from the initial surface $\Sigma'$ via the two-point kernel $\mbox{\boldmath$K$}(q,q')$ reads in the semiclassical approximation as
        \begin{equation}
        {\mbox{\boldmath$\varPsi$}}(q)
        =\int dq'\,{\mbox{\boldmath$K$}}(q,q')\,
        \delta(\chi(q',t'))\,
        \overrightarrow{J'}\,
        {\mbox{\boldmath$\varPsi$}}(q')
        +O(\hbar).                  \label{5.11b}
        \end{equation}

Relations (\ref{5.9})-(\ref{5.11b}) could have been exact beyond the semiclassical approximation if the first-class constraints were linear in momenta $p_i$. In this case the equations (\ref{5.11}) would have specified the derivatives of $\mbox{\boldmath$\varPsi$}(q)$ along gauge directions and the Faddeev-Popov determinant in the coordinate gauge would be just a $q$-dependent measure factor $J=J(q)$ independent of $p_i$. In quantum cosmology it is impossible, because the Hamiltonian constraint is quadratic in momenta, and $J(q,p)$ is a nonlinear function of momenta of power coinciding with the total number of Hamiltonian constraints (which is of course infinite in full gravity theory and formally equals the number of spatial points $\infty^3$).
\footnote{The semiclassical operator measure $\delta(\chi)\,\overrightarrow{\!\!J}+O(\hbar)$ in the physical inner product of Dirac wavefunctions in Eq.(\ref{5.11b}) can be promoted to the level of an exact concept by embedding the Dirac quantization into the BFV quantization \cite{BFV} in the extended phase space of all canonical pairs of ``matter" variables $q^i,p_i;N^\mu,\pi_\mu$ and pairs of Grassmann ghost variables $C^\mu,\bar{\cal P}_\mu;\bar C_\mu,{\cal P}^\mu$. This has been done within the concept of quantization on the so-called inner product spaces in \cite{BatalinMarnelius,WhyBFV}.}

Formal treatment of infinite number of degrees of freedom can be avoided in minisuperspace applications of quantum cosmology, when only one Hamiltonian constraint ($m=1$) survives in the finite dimensional phase space of the FRW metric and homogeneous matter fields. However, another problem still remains with the boundary value problem (\ref{5.11})-(\ref{5.11a}). The Wheeler-DeWitt equation (\ref{5.11}) is at least quadratic in derivatives $\partial/i\partial q$ and requires two initial conditions on $\Sigma$ -- the value of $\mbox{\boldmath$\varPsi$}(q)$ and its first order derivative, so that the number of solutions is at least doubled as compared to the reduced phase space dynamics. This ``positive-negative" frequency doubling serves as a source of a rather speculative third quantization concept \cite{third}, which represents the attempt to go beyond physical phase space reduction.

We will, however, remain within physical reduction concept which consists in lifting the physical wavefunction $\varPsi_{\rm phys}(t,\xi)$ to the level of the wavefunction in superspace $\mbox{\boldmath$\varPsi$}(q)$. The latter of course satisfies the Wheeler-DeWitt boundary value problem (\ref{5.11})-(\ref{5.11a}) but incorporates only the physical wavefunction information. In other words, a formal boundary value problem (\ref{5.11})-(\ref{5.11a}) contains many more solutions than the physically relevant ones which are encoded in the Schr\"odinger equation (\ref{SchroedEq}) of the reduced phase space quantization. Thus our task will be finding the selection rules for the solutions of the Wheeler-DeWitt equation appropriate for the physical setting in the reduced phase space.

\subsection{Basis of classical constraints and their operator realization}
Classical theory is of course invariant under the change of the basis of constraints with any invertible matrix $\varOmega^\nu_\mu=\varOmega^\nu_\mu(q,p)$,
    \begin{equation}
    H_\mu\to H'_\mu=
    \varOmega^\nu_\mu H_\nu, \label{basischange}
    \end{equation}
and under canonical transformations of phase space variables
    \begin{eqnarray}
    &&(q,p)\to (\tilde q,\tilde p), \label{cantran0}\\
    &&p_i\,dq^i-\tilde p_i\,
    d\tilde q^i=dF(q,\tilde q).     \label{cantran5}
    \end{eqnarray}
Here $F(q,\tilde q)$ is a generating function of this canonical transform, relating old and new symplectic forms. One should expect that at the quantum level this invariance should hold as unitary equivalence of Dirac quantization schemes in different constraint bases and different canonical parameterizations. At least in the semiclassical approximation (which includes one-loop prefactors) this issue formally has affirmative resolution. Interestingly, it is associated with the problem of operator realization of quantum constraints $\hat H_\mu$ which should satisfy a closed commutator algebra (\ref{1.6}).

As shown in \cite{BarvU,BKr,geom} in this approximation the quantum Dirac constraints are given by the Weyl ordering of their classical expressions
    \begin{equation}
    \hat H_\mu=N_{\rm W} H_\mu(\hat q,\hat p)
    +O(\hbar^2).                  \label{Weylorder}
    \end{equation}
Remarkably, this prescription holds in any basis of classical constraints. Under the basis change (\ref{basischange}) the formalism undergoes unitary transformation to new wavefunctions ${\mbox{\boldmath$\varPsi$}}'(q)$ and new operators $\hat H'_\mu$, $\hat H'_\mu{\mbox{\boldmath$\varPsi$}}'(q)=0$,
    \begin{eqnarray}
    &&\!\!\!\!{\mbox{\boldmath$\varPsi$}}(q)
    \to{\mbox{\boldmath$\varPsi$}}'(q)
    =\big({\rm det}\,
    \hat\varOmega^\nu_\mu\big)^{-1/2}\,
    {\mbox{\boldmath$\varPsi$}}(q)
    +O(\hbar),                    \label{tildepsi}\\
    &&\!\!\!\!\hat H_\mu\to\hat H'_\mu=\big({\rm det}\,
    \hat\varOmega^\alpha_\beta\big)^{-1/2}\,
    \hat\varOmega^\nu_\mu\,\hat H_\nu\,\big({\rm det}\,
    \hat\varOmega^\alpha_\beta\big)^{1/2}\nonumber\\
    &&\qquad\qquad\qquad\qquad\qquad\qquad\qquad\quad
    +O(\hbar^2).                           \label{HtoH}
    \end{eqnarray}
In other words, Dirac wavefunction ${\mbox{\boldmath$\varPsi$}}(q)$ is a density of minus one half weight in the space of gauge indices \cite{geom}.\footnote{$O(\hbar^2)$ and $O(\hbar)$ in (\ref{Weylorder}) and (\ref{tildepsi}) signify the same one-loop approximation because the semiclassical expansion for a quantum state begins with $1/\hbar$-order.}

Similarly, the classical canonical transformation (\ref{cantran0}) induces the unitary transformation (see derivation in Appendix A for a generic quantum system),
    \begin{eqnarray}
    &&{\mbox{\boldmath$\varPsi$}}(q)\to \tilde{\mbox{\boldmath$\varPsi$}}(\tilde q),\quad \hat{\widetilde H}_\mu\tilde{\mbox{\boldmath$\varPsi$}}(\tilde q)=0,\\
    &&{\mbox{\boldmath$\varPsi$}}(q)=
    \int d\tilde q\,
    \left|\,{\rm det}\,\frac1{2\pi\hbar}
        \frac{\partial^2F(q,\tilde q)}{\partial q^i\;\partial \tilde q^k}\,\right|^{1/2}\nonumber\\
        &&\qquad\qquad\quad
        \times\exp\left(\frac{i}{\hbar}\,F(q,\tilde q)\right)
        \tilde{\mbox{\boldmath$\varPsi$}}(\tilde q)
        +O(\hbar),                           \label{Fourier00}
        \end{eqnarray}
which was checked in \cite{geom} at least for contact transformations, $\tilde q=\tilde q(q)$, under which (\ref{Fourier00}) reduces to the transformation law of the weight 1/2 density
    \begin{eqnarray}
    &&{\mbox{\boldmath$\varPsi$}}(q)=
    \left|\,{\rm det}\,
        \frac{\partial\tilde q}{\partial q}\, \right|^{1/2}
        \tilde{\mbox{\boldmath$\varPsi$}}
        (\tilde q(q))+O(\hbar).        \label{Fourier01}
        \end{eqnarray}

These unitary equivalence transformations will be important in what follows, because we will have to go to another convenient canonical parametrization of the theory and also pick up a special normalization of the Hamiltonian constraint. Only their consistent treatment will guarantee correct transition between the physical and Wheeler-DeWitt wavefunctions of the theory.

\section{Minisuperspace models: from the physical sector to the Wheeler-DeWitt wavefunction}
The goals formulated in the end of Sect.II.B can be explicitly implemented in case of minisuperspace quantum cosmology with one Hamiltonian constraint. For the index $\mu$ taking in minisuperspace case only one value,
    \begin{equation}
    H_\mu(q,p)\equiv H(q,p),\quad \chi^\mu(q,p,t)\equiv\chi(q,p,t),
    \end{equation}
the gauge condition can be rewritten as expressing explicitly $t$ as a function on the phase space of $q^i$ and $p_i$,
    \begin{equation}
    \chi(q,p,t)=0\quad
    \Rightarrow\quad \chi(q,p,t)
    \equiv T(q,p)-t=0,              \label{Time}
    \end{equation}
so that the Faddeev-Popov determinant and the relevant lapse function (\ref{5.2})-(\ref{lapse0}) read
    \begin{equation}
    J=\{T,H\},\quad N=\frac1{J}.  \label{JvN}
    \end{equation}

The critical point of the physical reduction is the nondegeneracy of $J$ over the entire phase space. Degeneration of $J$ to zero at certain points in phase space implies the breakdown of the physical reduction known in the context of Yang-Mills type gauge theories as Gribov copies problem \cite{Gribov}. This problem arises when the surface of gauge conditions is not transversal everywhere to the orbits of gauge transformations and does not pick up a single representative of each class of gauge equivalent configurations. In gravity theory, and specifically in minisuperspace quantum cosmology, this problem manifests itself in the fact that the time variable $T(q,p)$ is not a monotonically growing function along all possible (on shell and off shell) histories. If $J$ changes sign, then according to (\ref{JvN}) $N$ changes sign too, and according to the geometrical interpretation of the lapse function the spacelike hypersurface of constant time $t$ with growing $t$ starts movin!
 g in spacetime back to the past. Therefore a meaningful physical reduction makes sense only in domains where $J$ does not change sign. In classical theory this problem is usually circumvented by such a choice of time which is monotonic at the classical history. The same applies to semiclassical quantization which probes only infinitesimal neighborhood of the latter. However, even in this simplified case the requirement of monotonic $T$ imposes serious restrictions on its choice as a function of phase space variables.

The role of the coordinate $q$ in minisuperspace models is basically played by the cosmological scale factor $a=e^\alpha$, so that
    \begin{equation}
    q^i,p_i=\alpha,p_\alpha;\;\xi,\pi,
    \end{equation}
where $\xi,\pi$ are actually the matter degrees of freedom other than $\alpha,p_\alpha$. This immediately forbids the choice of the coordinate gauge or time $T=T(\alpha)$ (intrinsic metric time) in problems with bouncing cosmology when the evolution of $a$ undergoes a bounce from some of its maximum or minimum values. More generally, the coordinate gauge conditions $\chi(q,t)$ are forbidden even semiclassically, because their $J= \{\chi,H\}$ vanishes on caustics in multidimensional superspace \cite{UFN} ($\dot a=0$ is the simplest degenerate case of the caustic in one-dimensional superspace). Using classical theory as a guide in the search for admissible gauges we will chose such $T(q,p)$ which monotonically grows along the classical trajectory, and in the bounce cosmology it immediately leads to $p$-dependent choice -- the so-called extrinsic time $T(\alpha,p_\alpha)$ \cite{extrinsic}.

This transition to momentum-dependent gauges implies that the quantum reduction (\ref{5.11a}) no longer applies directly. What we need is, first, to make a canonical transformation of the type (\ref{cantran0}),
    \begin{eqnarray}
    &&\alpha,p_\alpha\;\to\;
    T=T(\alpha,p_\alpha),\quad p_T=p_T(\alpha,p_\alpha),\nonumber \\
    && p_\alpha d\alpha-p_T dT=dF(\alpha,T),   \label{canonical}
    \end{eqnarray}
with the relevant generating function $F(\alpha,T)$. Second, we make classical and quantum reduction in terms of new phase space variables in the gauge (\ref{Time}). In these variables the Hamiltonian constraint and the Wheeler-DeWitt equation correspondingly read as
    \begin{eqnarray}
    &&H\equiv H(T,p_T;\xi,\pi)=0,   \label{HinT}\\
    &&\hat H\Big(T,\frac\partial{i\partial T};\hat\xi,\hat\pi\Big)\,
    |\,\widetilde{\mbox{\boldmath$\varPsi$}}(T)\,\rangle=0.
    \end{eqnarray}
The ket vector notation here refers to the state in the Hilbert space of matter operators $(\hat\xi,\hat\pi)$ and tilde labels the Wheeler-DeWitt wavefunction in the representation of the variable $T$. Quantum reduction to the physical sector involves the Faddeev-Popov ``determinant" and its operator realization
    \begin{eqnarray}
    &&J(T,p_T;\xi,\pi)=
    \frac{\partial H}{\partial p_T},\\
    &&\overrightarrow{\!\!J}=
    J\Big(T,\frac\partial{i\partial T};
    \hat\xi,\hat\pi\Big).           \label{JinT}
    \end{eqnarray}
The resulting physical wavefunction $|\,\varPsi_{\rm phys}(t)\,\rangle$ satisfies the Schr\"odinger equation (\ref{SchroedEq}).

In view of minisuperspace nature of the system the embedding (\ref{5.4}), $q=e(\xi,t)$, of the physical space into superspace of $q=(T,\xi)$ is in fact a one to one map between $q$ and the arguments $(t,\xi)$ of $|\,\varPsi_{\rm phys}(t)\,\rangle=\varPsi_{\rm phys}(t,\xi)$. Therefore Eq.(\ref{5.11a}) can be reversed to raise the physical quantum state to the level of the Wheeler-DeWitt wavefunction in superspace of the $T$-variable.
Because of $dq\,\delta(\chi(q,t))\equiv dT\,\delta(T-t)\, d\xi=d\xi$, cf. (\ref{M}), the integration measure in (\ref{5.11a}) $M=1$,  and this equation can be rewritten as\footnote{Here and in what follows we normalize for brevity $\hbar$ to unity and only label the commutator terms disregarded in the semiclassical approximation as $O(\hbar)$ or $O(\hbar^2)$.}
        \begin{eqnarray}
        |\,\widetilde{\mbox{\boldmath$\varPsi$}}(T)\,\rangle=
        \frac1{\big(\overrightarrow{\!\!J}\,\big)^{1/2}}\,
        |\,\varPsi_{\rm phys}(T)\,\rangle
        +O(\hbar).                              \label{5.11c}
        \end{eqnarray} 
The invertibility of $J$ semiclassically guarantees the invertibility of the corresponding operator coefficient acting on the $T$ argument of $|\,\varPsi_{\rm phys}(T)\,\rangle$.

Similarly to (\ref{Fourier00}) the canonical transformation (\ref{canonical}) implies at the quantum level the unitary transformation to the $\alpha$-representation from that of $T$
        \begin{eqnarray}
        |\,{\mbox{\boldmath$\varPsi$}}(\alpha)\,\rangle=
        \int_{-\infty}^{\infty} dT\,\langle\,\alpha\,|\,T\,\rangle\,
        |\,\widetilde{\mbox{\boldmath$\varPsi$}}(T)\,
        \rangle,          \label{genFtrans1}
        \end{eqnarray}
where
        \begin{equation}
        \langle\,\alpha\,|\,T\,\rangle=
        \left|\frac1{2\pi\hbar} \frac{\partial^2F(\alpha,T)}{\partial\alpha\;\partial T}\right|^{1/2}e^{\frac{i}\hbar\,F(\alpha,T)}
        +O(\hbar),   \label{genFtrans2}
        \end{equation}
whence
        \begin{eqnarray}
        &&|\,{\mbox{\boldmath$\varPsi$}}(\alpha)\,\rangle=
        \int_{-\infty}^{\infty} dT\,
        \left|\frac1{2\pi\hbar} \frac{\partial^2F(\alpha,T)}{\partial\alpha\;\partial T}\right|^{1/2}e^{(i/\hbar)\,F(\alpha,T)}\nonumber\\
        &&\qquad\qquad\quad\times
        \frac1{\big(\overrightarrow{J}\,\big)^{1/2}}\,
        |\,\varPsi_{\rm phys}(T)\,
        \rangle+O(\hbar).             \label{Fourier0}
        \end{eqnarray}
Note that this relation is nonlocal in time -- only the knowledge of the entire Schr\"odinger evolution of $|\,\varPsi_{\rm phys}(t)\,\rangle$ allows one to recover the Wheeler-DeWitt wavefunction in superspace of $q$.

Below we demonstrate how this equation works in several simple but essentially nonlinear minisuperspace models.
What will be important for us is whether the resulting wavefunction is either exponentially suppressed or infinitely oscillating at superspace boundaries $\alpha=-\infty$ and $\alpha=\infty$ (similarly for $T$ and $\hat p_T$ at the boundaries $T=\pm\infty$). This would guarantee Hermiticity of the phase space operators -- necessary property of the Dirac quantization and Wheeler-DeWitt equation formalism. Simultaneously this would provide us with the selection rules for physically motivated solutions of the Wheeler-DeWitt equation, which remove one half of the full set of their positive and negative frequency solutions. This restores the balance between the number of these solutions and the number of physical data in the reduced phase space quantization -- a single physical wavefunction $|\,\varPsi_{\rm phys}(t)\,\rangle$ at the initial Cauchy surface.

\section{Flat FRW model with a homogeneous scalar field}
We consider a flat Friedmann universe with the metric
    \begin{equation}
    ds^2 = N^2(t) dt^2
    - e^{2\alpha(t)}d{\bf x}^2, \label{Fried}
    \end{equation}
and a spatially homogeneous scalar field $\phi(t)$. Here
$N(t)$ is the lapse function and $e^{\alpha(t)}$ is the cosmological scale factor. The range of the minisuperspace variable $\alpha$
    \begin{equation}
    -\infty<\alpha<\infty              \label{range}
    \end{equation}
covers all values of the scale factor from singularity to infinite expansion.

With the normalization of the gravitational constant $1/16\pi G=3/4$ the action of this model
    \begin{equation}
    S = \int dx \sqrt{-g}\left(-\frac{R}{16\pi G} + \frac12g^{\mu\nu}\phi_{,\mu}\phi_{,\nu}
    - V(\phi)\right).                      \label{action}
    \end{equation}
gives rise to the minisuperspace Lagrangian and the Hamiltonian constraint
    \begin{eqnarray}
    &&L = -9e^{3\alpha}\frac{\dot\alpha^2}{2N}
    +e^{3\alpha}\frac{\dot{\phi}^2}{2N}
    -Ne^{3\alpha}V(\phi),             \label{Lagrange1}\\
    &&H=e^{-3\alpha}\left(-\frac1{18}\,p_{\alpha}^2 + \frac12\,p_{\phi}^2\right)
    + V(\phi)\,e^{3\alpha},     \label{H0}
    \end{eqnarray}
in terms of the canonical momenta for $\alpha$ and $\phi$
    \begin{equation}
    p_{\alpha} = -9\,\frac{\dot{\alpha}
    e^{3\alpha}}{N},\quad
    p_{\phi} = \frac{\dot{\phi}
    e^{3\alpha}}{N}.             \label{momentum2}
    \end{equation}

According to discussion in Sect.II.D there is a freedom in operator realization of this constraint associated with its overall normalization, $H\to H'=\varOmega\,H$ (choice of the constraint basis (\ref{basischange})). This freedom allows us to simplify the operator realization of $H'$. Multiplication of (\ref{H0}) with $\varOmega\equiv e^{3\alpha}$ converts the constraint into the form
    \begin{equation}
    H'=-\frac1{18}\,p_{\alpha}^2 + \frac12\,p_{\phi}^2
    + V(\phi)\,e^{6\alpha}.         \label{super}
    \end{equation}
The advantage of this normalization is that the kinetic term of $H$ is independent of minisuperspace coordinates, so that the Weyl ordering in the operator realization (\ref{Weylorder}) is trivial. In the coordinate representation it reduces to the replacement of momenta by partial derivatives
    \begin{equation}
    p_{\alpha} = -i\frac{\partial}{\partial \alpha},\quad
    p_{\phi} = -i\frac{\partial}{\partial \phi},
    \label{momentum4}
    \end{equation}
and the minisuperspace operator of the Wheeler-DeWitt equation for the cosmological wave function ${\mbox{\boldmath$\varPsi$}}(\alpha,\phi)$
    \begin{equation}
    \hat H'\,{\mbox{\boldmath$\varPsi$}}
    (\alpha,\phi)=0                              \label{WdW}
    \end{equation}
takes the form
    \begin{equation}
    \hat H'=\frac1{18} \frac{\partial^2}{\partial \alpha^2} - \frac12\frac{\partial^2}{\partial \phi^2}+ V(\phi)e^{6\alpha}. \label{WdW1}
    \end{equation}

\section{The case of a negative constant potential}
Variational equations for (\ref{Lagrange1}) in the gauge $N =1$ (the corresponding cosmic time we will denote by $\tau$) read
    \begin{eqnarray}
    &&\frac92\,h^2 =
    \frac{\dot{\phi}^2}{2} + V(\phi),        \label{Fried1}\\
    &&\ddot{\phi} + 3h\dot{\phi}
    + \frac{dV}{d\phi} = 0,                   \label{KG}
    \end{eqnarray}
where $h$ is the Hubble parameter
    \begin{equation}
    h \equiv \dot\alpha,
    \label{Hubble}
    \end{equation}
and unusual coefficient of $h^2$ in (\ref{Fried1}) is in fact $3/8\pi G=9/2$ in the chosen normalization of the gravitational constant.

These equations essentially simplify for a constant negative potential -- negative cosmological constant,
    \begin{equation}
    V(\phi) = -V_0,         \label{poten}
    \end{equation}
when the scalar field $\phi$ becomes a cyclic variable with a conserved momentum $p_{\phi}$. Then the equation for $\dot\phi$ in terms of $p_{\phi}$
    \begin{equation}
    \dot{\phi} = e^{-3\alpha}p_{\phi},   \label{velocity3}
    \end{equation}
when substituted in the Friedmann equation
(\ref{Fried1}) yields
    \begin{eqnarray}
    \frac92\,\dot\alpha^2 =
    \frac12\,p_{\phi}^2\,e^{-6\alpha} - V_0.  \label{Fried3}
    \end{eqnarray}
Integrating this equation one gets
    \begin{eqnarray}
    &&e^{\alpha(\tau)} = \left(|p_{\phi}|\,
    \frac{\sin \sqrt{2V_0}\tau}
    {\sqrt{2V_0}}\right)^{1/3},           \label{radius}\\
    &&h(\tau) = \frac{\sqrt{2V_0}}
    3\cot \sqrt{2V_0}\tau.                  \label{Hubble1}
    \end{eqnarray}
Thus, the Universe begins its evolution at $\tau = 0$ (Big Bang cosmological singularity), reaches the point of the maximal expansion at $\tau = \pi/2\sqrt{2V_0}$ when $h = 0$ and begins contracting to the Big Crunch singularity at $\tau = \pi/\sqrt{2V_0}$. During this evolution the
Hubble parameter is monotonously decreasing from $+\infty$ to $-\infty$. Below we consider quantum cosmology of this model. Note that the range of cosmic time is finite, $0<\tau<\pi/\sqrt{2V_0}$.

\subsection{General solution of the Wheeler-DeWitt equation}
With a constant potential (\ref{poten}) in view of cyclic nature of $\phi$ the wavefunction ${\mbox{\boldmath$\varPsi$}}(\alpha,\phi)$ is easily represented by its Fourier transform
        \begin{eqnarray}
        {\mbox{\boldmath$\varPsi$}}(\alpha,\phi)=
        \int_{-\infty}^{\infty} dp_\phi\,\varPsi(\alpha,p_\phi)\,
        e^{ip_\phi\phi}.
        \end{eqnarray}
The wavefunction in the momentum representation $\varPsi(\alpha,p_\phi)$ then satisfies the Wheeler-DeWitt equation
    \begin{equation}
    \left(\frac19\,\frac{\partial^2}{\partial \alpha^2}
    + p_{\phi}^2-2V_0 e^{6\alpha}\right)
    \varPsi(\alpha,p_{\phi}) = 0.         \label{WdW2}
    \end{equation}
This equation has a general solution \cite{Olver}
    \begin{eqnarray}
    &&\varPsi(\alpha,p_{\phi}) = \psi_1(p_{\phi})\,I_{i|p_{\phi}|}
    \big(\sqrt{2V_0}\,e^{3\alpha}\big)\nonumber \\
    &&\qquad\qquad\quad+\psi_2(p_{\phi})\,
    K_{i|p_{\phi}|}
    \big(\sqrt{2V_0}\,e^{3\alpha}\big),  \label{WdW-sol}
    \end{eqnarray}
where $I_{\nu}(x)$ and $K_{\nu}(x)$ are the modified Bessel functions of the first and the second kind respectively, whereas  $\psi_1(p_{\phi})$ and $\psi_2(p_{\phi})$ are generic functions of the momentum $p_{\phi}$.

The two branches of the generic solution (\ref{WdW-sol}) are drastically different. Near the cosmological singularity, $\alpha\to-\infty$, in view of imaginary value of $\nu=i|p_\phi|$ they both represent plane waves $\sim e^{\pm 3\nu\alpha}=e^{\pm 3i|p_\phi|\alpha}$. However, for $x=e^{3\alpha}\to+\infty$ one of them is rapidly growing, $I_\nu(x)\propto e^x/\sqrt{x}=\exp(e^{3\alpha}-3\alpha/2)$, and another is exponentially decaying $K_\nu(x)\propto e^{-x}/\sqrt{x}$. Therefore, Hermiticity of canonical momenta $\hat p_\alpha=-i\partial/\partial\alpha$ and other operators with respect to the $L_2$ inner product on the range of $\alpha$ (\ref{range}) is possible only for quantum states represented by the second branch of the solution (\ref{WdW-sol}). These Hermiticity properties are very important in the Dirac quantization scheme and in even more general BRST/BFV quantization scheme \cite{BFV,BarvU,WhyBFV}.\footnote{BRST quantization has as one of its basic ingredients, the (!
 unphysical) inner product of $L_2$-type in the bosonic sector of phase space and Berezin integration inner product in the sector of its Grassmann ghost variables.} Violation of these properties leads to inconsistency of the formalism. Therefore, consistency of Dirac quantization should serve as a selection rule which retains only the MacDonald function branch $K_{i|p_{\phi}|}
\big(\sqrt{2V_0}\,e^{3\alpha}\big)$ of (\ref{WdW-sol}). Below we show that in this particular model the same selection rule follows from quantization in the physical sector. 

\subsection{Physical wavefunction}
From the discussion of Sect.II.C (see Eqs. (\ref{Time})-(\ref{JvN})) we remember that the gauge condition in the reduction to the physical sector can be cast into the form
    \begin{equation}
    \chi(T,t) = T-t, \label{gauge}
    \end{equation}
explicitly depending on time $t$, and $T=T(\alpha,p_\alpha)$ is a new canonical variable, depending on the old phase space variables. For consistency of physical reduction, such a variable should monotonously grow with time at least on classical solutions of the model. For models with the cosmological expansion followed by contraction, the variables depending only on the scale factor $a$ (or its logarithm $\alpha$) are not monotonous. Thus, $T(\alpha,p_\alpha)$ should involve the canonical momentum, and it is called extrinsic time  \cite{extrinsic} (involving extrinsic curvature rather than intrinsic geometry of a spatial surface in spacetime). A possible choice is the following function
    \begin{equation}
    T(\alpha,p_\alpha) =  \frac1{3\sqrt2}\, p_{\alpha}e^{-3\alpha},   \label{choice}
    \end{equation}
which is proportional to the Hubble parameter\footnote{Note that the extrinsic time introduced in minisuperspace by Eqs. (\ref{gauge})-(\ref{choice}) coincides un to a numerical factor with the York extrinsic time  introduced in \cite{extrinsic} for an arbitrary manifold $\tau \equiv \frac23\gamma^{-1/2}\pi$, $\pi = \gamma^{ab}\pi_{ab}$, where $\gamma_{ab}$ is a spatial metric and $\pi^{ab}$ is its conjugated momenta.} and therefore classically has infinite range, $-\infty<T<\infty$ (cf. beginning of Sect.V). The conjugated momentum for this variable is
    \begin{equation}
    p_{T} = -\sqrt2\, e^{3\alpha}. \label{momentum5}
    \end{equation}
so that the inverse transform from $(T,P_T)$ to $(\alpha,p_\alpha)$ reads
    \begin{eqnarray}
    &&e^{3\alpha} = -\frac{p_T}{\sqrt2},  \label{inv}\\
    &&p_{\alpha} = -3p_T T.  \label{inv1}
    \end{eqnarray}

The constraint (\ref{super}) in terms of new variables
    \begin{equation}
    H'(T,p_T,p_\phi)\equiv-\frac12\,p_T^2 (T^2 + V_0)+\frac12 \,p_{\phi}^2=0
    \label{constraint1}
    \end{equation}
has a solution for $p_T$
    \begin{equation}
    p_T = -\frac{|p_{\phi}|}{\sqrt{T^2+V_0}},
    \end{equation}
where a particular sign of the square root is chosen in accordance with the geometrical meaning of the momentum $p_T$ (minus the 3-dimensional volume of the cosmological model). Thus, the physical Hamiltonian in the gauge (\ref{gauge}) reads (cf. Eq. (\ref{5.8a}))
    \begin{equation}
    H_{\rm phys}(p_\phi,t) = -p_T\,\big|_{\,T=t} = \frac{|p_{\phi}|}{\sqrt{t^2+V_0}}.
    \label{Hamilton-phys}
    \end{equation}

The corresponding Schr\"odinger equation for the physical wave function
\begin{equation}
    i\,\frac{\partial \varPsi_{\rm phys}(t,p_\phi)}{\partial t} = H_{\rm phys}(p_{\phi},t)\,
    \varPsi_{\rm phys}(t,p_\phi).      \label{Schrodinger}
    \end{equation}
immediately gives the time evolution
    \begin{eqnarray}
    &&\varPsi_{\rm phys}(\tau,p_\phi) = \psi_0(p_{\phi}) \exp\left(-i|p_\phi|\,{\rm arcsinh} \frac{t}{\sqrt{V_0}}\right)\nonumber\\
    &&=\psi_0(p_{\phi}) \left(\frac{\sqrt{V_0}}{t+\sqrt{t^2+V_0}}
    \right)^{i|p_\phi|},             \label{Schrodinger1}
    \end{eqnarray}
where $\psi_0(p_{\phi})$ is an arbitrary function of the momentum $p_{\phi}$ -- the initial data  for (\ref{Schrodinger}) at $t=0$. Note that the time $t$ in contrast to the cosmic time $\tau$ has infinite range $-\infty<t<\infty$, and Eq.(\ref{Schrodinger}) propagates the data from $t=0$ both forward and backward in time.

\subsection{From the physical wave function to the solution of the Wheeler-DeWitt equation}
To lift the physical wavefunction to the level of the Wheeler -DeWitt wavefunction according to (\ref{5.11c}) we have to act with the inverse square root of the operator version of the Faddeev-Popov operator. In the gauge (\ref{gauge}) it reads
    \begin{equation}
    J=\{T, H'\}= -p_T(T^2 + V_0),       \label{JP1}
    \end{equation}
so that semiclassically
\begin{equation}
    \overrightarrow{\!\!J}=-(T^2 + V_0)\,\frac\partial{i\,\partial T},
    \end{equation}
Thus, with the same one-loop precision (disregarding the operator ordering in the above equation)
    \begin{equation}
    \tilde\varPsi(T,p_\phi) = \left(-\frac\partial{i\,\partial T}\right)^{-1/2}
    \frac{\varPsi_{\rm phys}(T,p_\phi) }
    {\sqrt{T^2+V_0}}+O(\hbar).           \label{JP2}
    \end{equation}

Now by the generalized Fourier transform (\ref{genFtrans1})-(\ref{genFtrans2}) we have to go from the $T$-representation to $\alpha$-representation. The classical generating function $F(\alpha,T)$ of this transform equals
    \begin{eqnarray}
    &&F(\alpha,T)=\sqrt2\,e^{3\alpha}T,\\
    &&p_\alpha d\alpha-p_T dT=dF(\alpha,T).
    \end{eqnarray}
Therefore the kernel of the unitary transformation (\ref{genFtrans2}) reads
    \begin{equation}
    \langle\,\alpha\,|\,T\,\rangle=
    \sqrt{\frac3{\pi\sqrt2}}\,e^{3\alpha/2+i\sqrt2\,e^{3\alpha}T}
    +O(\hbar),
    \end{equation}
and
    \begin{eqnarray}
    &&\varPsi(\alpha,p_{\phi})=\int\limits_{-\infty}^{\infty}dT\,
    \langle\,\alpha\,|\,T\,\rangle\,
    \tilde\varPsi(T,p_\phi)\nonumber\\
    &&\qquad\quad=\sqrt{\frac3{\pi\sqrt2}}\; e^{3\alpha/2}\int\limits_{-\infty}^{\infty}dT\,
    e^{i\sqrt2\,e^{3\alpha}T}\nonumber\\
    &&\qquad\quad\times\left(-\frac\partial{i\,\partial T}\right)^{-1/2}
    \frac{\varPsi_{\rm phys}(T,p_\phi) }
    {\sqrt{T^2+V_0}}+O(\hbar).        \label{X}
    \end{eqnarray}
``Integrating by parts" the time derivative in the nonlocal operator $(-\partial/i\partial T)^{1/2}$ -- that is acting by $(\partial/i\partial T)^{1/2}$ to the left on the exponential function of $T$ (which is justified by rapid oscillations of the integrand, cf. Eq.(\ref{Schrodinger1}), and decrease of its amplitude at $T\to\pm\infty$) -- we get
    \begin{equation}
    \varPsi(\alpha,p_{\phi})
    =\sqrt{\frac3{2\pi}}\,
    \int\limits_{-\infty}^{\infty}\frac{dT\,\varPsi_{\rm phys}(T,p_\phi)}{\sqrt{T^2+V_0}}\,
    e^{i\sqrt2\,e^{3\alpha}T}+O(\hbar).   \label{XX}
    \end{equation}
Substituting (\ref{Schrodinger1}) and using as a new integration variable
    \begin{equation}
    x \equiv  {\rm arcsinh} \frac{T}{\sqrt{V_0}},
    \label{new-var}
    \end{equation}
we have
    \begin{equation}
    \varPsi(\alpha,p_{\phi}) =\sqrt{\frac3{2\pi}}\, \psi_0(p_{\phi})\int\limits_{-\infty}^{\infty}dx\,
    e^{-i\,|p_{\phi}|\,x
    +i\,e^{3\alpha}\sqrt{2V_0}\,\sinh x}.   \label{Fourier1}
    \end{equation}
Then, in virtue of the formula for the MacDonald function
    \begin{equation}
    K_{\nu}(x)=\frac12\, e^{\frac12 \nu\pi i}
    \int\limits_{-\infty}^{\infty}dt\,
    e^{ix\sinh t -\nu t},                 \label{Watson2}
    \end{equation}
which is valid for real positive $x$ and the order $\nu$ belonging to the interval $-1 < {\rm Re}(\nu) < 1$ (see Appendix B for derivation), the wavefunction eventually takes the form
    \begin{equation}
    \varPsi(\alpha,p_{\phi}) =\sqrt{\frac{6}{\pi}}\,\psi_0(p_{\phi})\,
    e^{- \frac12\pi|p_{\phi}|}
    K_{i|p_\phi|}
    \big(\sqrt{2V_0}e^{3\alpha}\big).          \label{M-K3}
    \end{equation}
This is just one branch of the general solution of the Wheeler-DeWitt equation (\ref{WdW-sol}) with\footnote{Note that the normalization of $H$ chosen in (\ref{super}) should be kept fixed throughout the calculation leading to (\ref{M-K3}). In particular it leads to the concrete normalization of $J\sim p_T$ in (\ref{JP1}), origin of $(-\partial/i\partial T)^{-1/2}$ in (\ref{X}) and cancelation of $e^{3\alpha/2}$ in (\ref{XX}). Without this the resulting wavefunction would not satisfy the Wheeler-DeWitt equation in the operator realization (\ref{WdW2}).}
    \begin{eqnarray}
    &&{\psi}_1(p_\phi) = 0,  \label{restr}\\
    &&{\psi}_2(p_\phi) =
    \sqrt{\frac{6}{\pi}}\,\psi_0(p_{\phi})
    \exp\left(-\frac{\pi|p_{\phi}|}2\right).            \label{restr1}
    \end{eqnarray}
Thus, we have only one independent function in the solution of the Wheeler-DeWitt equation. Remarkably, this is exactly the MacDonald branch of (\ref{WdW-sol}) which guarantees Hermiticity of the momentum operator $\hat p_\alpha=\partial/i\partial\alpha$ and makes Dirac quantization scheme consistent.

\subsection{Cosmic time}

Physical reduction can also be done in the cosmic time gauge. To find the extrinsic time variable $\tilde T(q,p)$ that would generate cosmic time with $N=1$ in phase space, one should solve the differential equation for $\tilde T$ in partial derivatives, $\{\tilde T,H'\}=e^{3\alpha}= -p_T/\sqrt2$.\footnote{The normalization of the constraint $H'=e^{3\alpha}H$ implies rescaling the lapse function -- the Lagrangian multiplier for the constraint -- $N'=e^{-3\alpha}N$, so that in view of (\ref{JvN}) the cosmic time corresponds to $\{\tilde T,H'\}=1/N'=e^{3\alpha}$.} This solution $\tilde T$ turns out to be related to $T$ by the contact transformation
    \begin{equation}
    \tilde T(\alpha,p_\alpha) = \frac{1}{\sqrt{2V_0}}\,{\rm arccot}\left(-\frac{T(\alpha,p_\alpha)}{\sqrt{V_0}}\right), \label{T-n}
    \end{equation}
where $T(\alpha,p_\alpha)$ is defined by Eq.(\ref{choice}). It is instructive to demonstrate that starting with the physical wavefunction $\tilde\varPsi_{\rm phys}(\tau,p_\phi)$ built in the gauge $\tilde\chi\equiv\tilde T-\tau=0$ one again comes to the solution of the Wheeler-DeWitt equation (\ref{WdW2}).

Repeating the procedure of reduced phase space quantization in this gauge it is easy to see the set of relations between quantization schemes with $T$ and $\tilde T$ time variables
    \begin{eqnarray}
    &&\tilde J\equiv\{\tilde T,H'\}=\frac{\partial\tilde T}{\partial T}\,J,\\
    &&\overrightarrow{\!\!\tilde J}=\frac{\partial\tilde T}{\partial T}\,\overrightarrow{\!\!J},\\
    &&\tilde\varPsi_{\rm phys}(\tilde T,p_\phi)
    =\varPsi_{\rm phys}(T(\tilde T),p_\phi).
    \end{eqnarray}
There is also the relation between the generating functions of canonical transformation from $T$ to $\alpha$ and from $\tilde T$ to $\alpha$, $\tilde F(\alpha,\tilde T)=F(\alpha, T(\tilde T))$, so that
    \begin{equation}
    \frac{\partial^2\tilde F(\alpha,\tilde T)}{\partial\alpha\,\partial\tilde T}=
    \frac{\partial^2F(\alpha,T)}{\partial\alpha\,\partial T}\,
    \frac{\partial T}{\partial\tilde T}.
    \end{equation}
Using these relations in the tilde version of (\ref{Fourier0}) we see that, after the change of integration variable from $\tilde T$ to $T(\tilde T)$ all factors of $\partial\tilde T/\partial T$ cancel out, and it yields the same Wheeler-DeWitt wavefunction. This confirms the anticipated property of the formalism that quantization schemes in different gauges give rise to one and the same Wheeler-DeWitt wavefunction.

\section{The case of a vanishing potential}

Qualitatively different situation in the above model takes place in the case of a vanishing potential. Its classical evolution (\ref{radius}) in the limit $V_0\to 0$ becomes
    \begin{equation}
    e^\alpha=(|p_{\phi}|\,\tau)^{1/3},
    \quad  0<\tau<\infty, \label{radius1}
    \end{equation}
for the positive range of cosmic time and corresponds to cosmological expansion from singularity to infinite scale factor. With $V_0=0$ the relation (\ref{T-n}) between cosmic time $\tau$ and time variable $t$ reads
    \begin{equation}
    \tau=-\frac1{\sqrt2\, t}, \label{tau-t}
    \end{equation}
so that (\ref{radius1}) maps onto
    \begin{equation}
    e^\alpha=\left(-\frac{|p_{\phi}|}t\right)^{1/3},
    \quad -\infty<t<0, \label{radius10}
    \end{equation}
on a negative range of $t$. The contracting stage of the cosmological evolution can be described by opposite ranges of $\tau$ and $t$
    \begin{eqnarray}
    e^\alpha&=&(-|p_{\phi}|\,\tau)^{1/3}\nonumber\\
    &=&
    \left(\frac{|p_{\phi}|}t\right)^{1/3},\;
    0<t<\infty,\;-\infty<\tau<0.      \label{radius11}
    \end{eqnarray}

Both expansion and contraction can be unified as consecutive stages of a single evolution by gluing together semi-axes of $\tau$ or $t$
    \begin{equation}
    e^\alpha=(-|p_{\phi}|\,|\tau|)^{1/3}=
    \left(\frac{|p_{\phi}|}{|t|}\right)^{1/3},\quad
    -\infty<\tau,t<\infty.       \label{radius12}
    \end{equation}
With this identifications the transition through the point $\tau=0$ implies the ``bounce" at the singularity $e^\alpha=0$, whereas a similar transition through $t=0$ can be interpreted as a ``turning" point from expansion to contraction at infinite scale factor, $e^\alpha\to\infty$. This unification is not physically natural however, because this transition through $\tau=0$ and $t=0$ lacks continuity and violates matching physical data at these junction points. As we will see below, this is even more manifest within physical reduction both at classical and quantum levels.

At the quantum level the general solution of the Wheeler-DeWitt equation (\ref{WdW2}) for $V_0 = 0$ represents a pure plane wave
    \begin{equation}
    \varPsi(\alpha,p_{\phi}) = \psi_1(p_{\phi})\,e^{3i|p_{\phi}|\alpha} +
    \psi_2(p_{\phi})\,e^{-3i|p_{\phi}|\alpha}.\label{WdWmassless}
    \end{equation}
Therefore, in contrast to (\ref{WdW-sol}) Hermiticity of momentum operator does not impose restrictions on coefficient functions $\psi_{1,2}(p_\phi)$. However, physical reduction continues selecting only one independent branch of this general solution.

By choosing the range of time variable $T\leq 0$ or $T\geq 0$ one can restrict dynamics in the physical sector entirely to expanding or contracting phases of the evolution, the way it happens on solutions of equations of motion in classical theory. From the geometric meaning of $p_T$ as a negative quantity (\ref{momentum5}) it follows that it reads as a solution of the Hamiltonian constraint $p_T=-|p_\phi|/|T|$ for both signs of $T$, and the physical Hamiltonian $H_{\rm phys}=|p_\phi|/|t|$ at the quantum level gives
    \begin{equation}
    \varPsi_{\rm phys}^\pm(t,p_\phi)=
    \psi(p_\phi)\,\left(\frac{t_0}t\right)^{\pm i|p_\phi|}
    \end{equation}
respectively for contracting $(+)$ and expanding $(-)$ cases, for which $t$ and $t_0$ are correspondingly positive and negative. Here $t_0$ is the initial data moment of time when $\varPsi_{\rm phys}^\pm(t_0,p_\phi)=\psi(p_\phi)$, and it has the same sign as $t$. Then integration in Eq.(\ref{XX}) with $V_0=0$ respectively over positive and negative values of $T$ gives
    \begin{eqnarray}
    &&\varPsi^\pm(\alpha,p_{\phi})
    =\sqrt{\frac3{2\pi}}\,\psi(p_\phi)\,
    \left(\sqrt2|\,t_0\,|\right)^{\pm i|p_\phi|}\nonumber\\
    &&\qquad\qquad\qquad\!\times\, e^{\frac{\pi|p_\phi|}2}\varGamma\big(\mp i|p_\phi|\,\big)\,
    e^{\pm 3i\,|p_\phi|\alpha},
    \end{eqnarray}
which are of course the two branches of (\ref{WdWmassless}).\footnote{Note that normalizability of this wavefunction with respect to $L_2$ inner product in $p_\phi$-space is the same as that of $\psi(p_\phi)$, because $e^{\pi|p_\phi|/2}|\varGamma\big(\mp i|p_\phi|\,\big)|\sim\sqrt{2\pi/|p_\phi|}$, for $|p_\phi|\to\infty$.} Thus, the physical reduction separately for contracting and expanding cosmological models leads to selection of a relevant branch in the full Wheeler-DeWitt solution.

How natural is the unification of these two branches into a single picture mentioned above? In physical reduction this unification is possible only by the price of violating the geometrical meaning of $p_T$ as a negative quantity (\ref{momentum5}), because the requirement of analyticity demands a replacement $H_{\rm phys}=|p_\phi|/|t|\to H_{\rm phys}=|p_\phi|/t$. For the full time range, $-\infty<t<\infty$, this generates a physical wavefunction
    \begin{equation}
    \varPsi_{\rm phys}(t,p_\phi)=
    \psi(p_\phi)\,\left(\frac{t_0}t\right)^{i|p_\phi|}.
    \end{equation}
It has a branching point at $t=0$ and requires a prescription for analytical continuation either from $t>0$ to $t<0$ (``contracting" wavefunction) or from $t<0$ to $t>0$ (``expanding" wavefunction). Moreover, physical reduction at $T=0$ also becomes inconsistent because the main ingredient of this reduction, the Faddeev-Popov factor $J=-p_TT^2$, gets degenerate at $T=0$, and again a special prescription is needed how to detour this point by the path in the complex plane of $T$. Provided this is done, one can apply Eq.(\ref{XX}) with integration over a full range of $T$, $-\infty<T<\infty$, and acquire
    \begin{equation}
    \varPsi(\alpha,p_{\phi})=
    \left\{\begin{array}{l}  0,\\
    \\
     \big(1-e^{-2\pi|p_\phi|}\big)\,
     \varPsi^+(\alpha,p_{\phi}),
    \end{array}\right.                \label{betas}
    \end{equation}
where the first case corresponds to the analytic continuation of $\varPsi_{\rm phys}(t,p_\phi)$ from the positive values of $t$ to the upper shore of the branch cut at negative semi-axis of $t$, $\varPsi_{\rm phys}(-|t|,p_\phi)=\varPsi_{\rm phys}(e^{i\pi}|t|,p_\phi)$, and the second case corresponds to the lower shore of this cut. Similar expressions in terms of $\varPsi^-(\alpha,p_{\phi})$ can be obtained if we start with the physical ``expanding" wavefunction analytic near the negative semi-axis of $t$ and continue it to the branch cut along the positive semi-axis. In both cases only one branch of the general solution (\ref{WdWmassless}) $\varPsi^\pm(\alpha,p_{\phi})$ is generated and physical reduction leads to selection of Wheeler-DeWitt wavefunctions. Similar conclusions can be reached within cosmic time reduction with extrinsic time variable $\tilde T$.

Now it is hard to say how meaningful is this unification of expanding and contracting stages. This is, of course, a certain extension of the quantization concept in physical sector. Within the latter this unification seems as contrived as it is in classical theory. Classical solutions having no turning points at large values of the scale factor and no bounces close to singularities imply that quantum dynamics is also entirely restricted to either expansion or contraction, and both of them are related by time inversion.

\subsection{Intrinsic time}
Absence of turning points in classical dynamics implies that physical reduction can be done in the intrinsic time -- the situation when the time variable $T$ is chosen as a function of only 3-metric of the theory and does not involve its conjugated momenta. In minisuperspace context this means identifying $T$ with, say, $\alpha$ and imposing a simple gauge
    \begin{equation}
    \chi = \alpha-t.     \label{gauge-alpha}
    \end{equation}
Then the physical Hamiltonian equals $-p_{\alpha}$ and as a solution of the constraint in our simple model with a vanishing scalar potential $V_0=0$ reads
    \begin{equation}
    H_{phys} = -p_{\alpha} =-3\varepsilon\,|p_{\phi}|. \label{Ham-double}
    \end{equation}
In this case we do not have any reason to disregard any of the $\varepsilon=\pm 1$ sign factors which correspond respectively to expansion and contraction. Initial value data at a given $t$, that is at a given spacelike surface with the scale factor $e^{3t}$, includes the discrete degree of freedom $\varepsilon$ indicating the direction of evolution. The physical wavefunction becomes a two-component vector $\varPsi_{\rm phys}^\varepsilon(t,p_\phi)$ whose components evolve with time differently under the action of the Hamiltonian (\ref{Ham-double}). Lifting this state to the level of the Wheeler-DeWitt wavefunction implies the generalized Fourier transform, analogous to (\ref{X}) but apparently including summation over discreet values of $\varepsilon$. In this simple model with $V_0=0$ it trivially leads to the superposition (\ref{WdWmassless}) with absolutely independent functions $\psi_{1,2}(p_\phi)$.

Note that in the case of the vanishing potential the Hermiticity condition discussed above does not impose restrictions on the general solution of the Wheeler-DeWitt equation. The two terms in the general solution correspond respectively to expanding and contracting universes. When we construct the physical wave function, based on the intrinsic time choice we again obtain two physical wave functions, which can be lifted to the level of two branches of the general solution of the Wheeler-DeWitt equation.

However, if we consider instead the extrinsic time parameter, then the corresponding physical wave function, when lifted to the level of the solution of the Wheeler-DeWitt equation, contains only one independent solution of this equation. This is explained by the fact that introduction of the extrinsic time implies a unique evolution instead of two independent evolutions -- contraction and expansion.
Namely, we have expansion-infinity-contraction for the time parameter associated with the Hubble variable and contraction-singularity-expansion for the cosmic time parameter.

Selection of one branch of the Wheeler-DeWitt equation solution via physical reduction with external time might look unnatural. Such a selection is not enforced by Hermiticity requirement, but rather arises  as an artifact of unifying the expansion and contraction of the universe as stages of unique evolution, whereas physically in this model these stages are separated either by cosmological singularity or by a domain of asymptotically infinite size of the universe.
Thus, intrinsic time treatment and intrinsic time setting of Cauchy problem seems more natural here, because it yields two-component physical wavefunctions, which give rise to two independent branches of the Wheeler-DeWitt wavefunction, describing two different types of evolution in quantum ensemble -- expansion and contraction.

\section{Phantom scalar field with a positive constant potential}
In order to see that nontrivial selection of solutions of the Wheeler-DeWitt equation matches with Hermiticity requirements in Dirac quantization we consider another example -- a phantom scalar field with a positive constant potential. This field has a negative kinetic term,
    \begin{equation}
    S = \int dx \sqrt{-g}\left(-\frac{R}{16\pi G}- \frac12g^{\mu\nu}\phi_{,\mu}\phi_{,\nu}
    - V_0\right),V_0>0,                 \label{phatomaction}
    \end{equation}
and despite obvious violation of unitarity in scattering problems this model recently attracted a lot of interest in context of dark energy models \cite{phantom}. As we will see it also raises interesting issues of underbarrier semiclassical behavior and boundaries in cosmological minisuperspace. This happens because this model has a cosmological bounce at small values of the scale factor -- the situation similar to known prescriptions for the cosmological wave function of the universe, based on the ideas of Euclidean quantum gravity and quantum tunneling \cite{HH,Vil,tun1,tun-we}.

The Friedmann equations (\ref{Fried1})-(\ref{Fried3}) with inverted kinetic term of the scalar field and inverted constant potential ($p_\phi^2\to-p_\phi^2$, $V_0\to-V_0$) have a solution which in the cosmic time $\tau$ and the Hubble time $t$ of Sect.V read
    \begin{eqnarray}
    &&e^{3\alpha} =|p_{\phi}|\,
    \frac{\cosh\sqrt{2V_0}\tau}{\sqrt{2V_0}}
    =\frac{|p_\phi|}{\sqrt{2(V_0-t^2)}},\label{phantom2}\\
    &&h=\frac{\sqrt{2V_0}}3\,\tanh \left(\sqrt{2V_0}\tau\right)
    =\frac{\sqrt2}3\,t.        \label{phantom1}
    \end{eqnarray}
Time variables of this cosmological evolution from the moment of infinite size to the bounce and back to the moment of infinite expansion run in the ranges
    \begin{equation}
    -\infty<\tau<\infty,
    \quad -\sqrt{V_0}<t<\sqrt{V_0}. \label{t-phantom}
    \end{equation}
In contrast to the model of Sect.V here the range of Hubble  time is compact.

At the level of Dirac quantization the Wheeler-DeWitt equation here reads
    \begin{equation}
    \left(\frac19\,\frac{\partial^2}{\partial \alpha^2}-p_{\phi}^2  +2V_0\,e^{6\alpha}\right)
    \varPsi(\alpha,p_{\phi}\phi)=0          \label{WdW2-phan}
    \end{equation}
and has a general solution
    \begin{eqnarray}
    &&\varPsi(\alpha,p_{\phi}) = \psi_1(p_{\phi})\,J_{|p_\phi|}
    \big(\sqrt{2V_0}e^{3\alpha}\big)   \nonumber \\
    &&\qquad\qquad\quad
    +\psi_2(p_{\phi})\,J_{-|p_\phi|}
    \big(\sqrt{2V_0}e^{3\alpha}\big),  \label{WdW-sol-phan}
    \end{eqnarray}
where $J_{\nu}(x)$ are the Bessel functions. Their leading behavior at $x \to \infty$
    \begin{equation}
    J_{\nu} (x) \sim \sqrt{\frac{2}{\pi x}}\cos\left(x-\frac{\nu\pi}{2}
    -\frac{\pi}{4}\right),          \label{Bessel2-ph}
    \end{equation}
and therefore Hermiticity requirement at $\alpha \rightarrow \infty$ does not impose any restriction on the branches of (\ref{WdW-sol-phan}). However, at the cosmological singularity $\alpha \rightarrow -\infty$ the Bessel function of a negative order diverges as $x^{\nu}$, and the
requirement of Hermiticity selects only the first term of (\ref{WdW-sol-phan}).

Let us see if this selection also works if we start with quantization in the physical sector. Repeating the steps of Sect.V in the gauge $\chi = T-\tau=0$, where the Hubble time variable is chosen similarly to (\ref{choice}) (opposite sign is taken to match the start of contraction from infinity with the negative value $T=-\sqrt{V_0}$)
    \begin{equation}
    T = -\frac1{3\sqrt2}\,
    p_{\alpha}e^{-3\alpha},\quad
    p_T = \sqrt2\, e^{3\alpha}.  \label{bar1}
    \end{equation}
The Hamiltonian constraint, Faddeev-Popov factor and the physical Hamiltonian now read
    \begin{eqnarray}
    &&H'(T,p_T,p_\phi)=-\frac12\,
    p_T^2 (T^2-V_0)
    -\frac12 \,p_{\phi}^2,       \label{bar4}\\
    &&J = -p_T\,(V_0-T^2),       \label{FP-ph}\\
    &&H_{\rm phys} = -p_{\bar{T}}= -\frac{|p_{\phi}|}{\sqrt{V_0-t^2}},    \label{Ham-ph}
    \end{eqnarray}
and the relevant solution of the Schr\"odinger equation is
    \begin{equation}
    \varPsi_{\rm phys}(t,p_{\phi}) = \psi_0(p_{\phi})\exp\left(i\,|p_{\phi}|\,{\rm arcsin}\frac{t}{\sqrt{V_0}} \right), \label{Schrod-ph}
    \end{equation}
where of course $\psi_0(p_{\phi})$ is the physical data at $t=0$.

Similarly to (\ref{X})-(\ref{XX}) the transition from $\varPsi_{\rm phys}(t,p_{\phi})$ to $\varPsi(\alpha,p_{\phi})$ becomes
    \begin{equation}
    \varPsi(\alpha,p_{\phi})
    =\sqrt{\frac3{2\pi}}\,
    \int\limits_{-\sqrt{V_0}}^{\sqrt{V_0}}\frac{dT\,\varPsi_{\rm phys}(T,p_\phi)}{\sqrt{T^2-V_0}}\,
    e^{-i\sqrt2\,e^{3\alpha}T}+O(\hbar),   \label{XXXX}
    \end{equation}
where we retain the integration range (\ref{t-phantom}) where only the unitary evolution with a {\em real} ${\rm arcsin}(t/\sqrt{V_0})$ is possible. Introducing a new variable
    \begin{equation}
    \theta \equiv {\rm arcsin}\frac{T}{\sqrt{V_0}}.
    \label{theta}
    \end{equation}
we get
    \begin{eqnarray}
    &&\varPsi(\alpha,p_{\phi})
    =\sqrt{6\pi}\,
    \psi_{0}(p_{\phi})\,I(x,\nu),     \label{psi} \\
    &&I(x,\nu)\equiv\frac1{2\pi}
    \int\limits_{-\pi/2}^{\pi/2}
    d\theta\,e^{-ix\sin\theta +i\nu\theta}, \label{I}\\
    &&x\equiv\sqrt{2V_0}\,
    e^{3\alpha},\quad \nu\equiv|p_{\phi}|.  \label{xnu}
    \end{eqnarray}

\subsection{Underbarrier domains and minisuperspace boundaries}
Now we have to remember that our formalism of quantum physical reduction is known only semiclassically -- up to $O(\hbar)$ terms which extend beyond one-loop order (classical exponent and prefactor). So let us check consistency of the obtained result and, in particular, the integration range in (\ref{I}) within this approximation. It corresponds to the limit when both $x$ and $\nu$ are large,
    \begin{equation}
    x,\,\nu=O\left(\frac1\hbar\right)\to\infty.
    \label{classical}
    \end{equation}
Classically allowed domain is defined by the overbarrier range of these parameters (\ref{xnu})
    \begin{equation}
    x>\nu,           \label{overbarrier}
    \end{equation}
where the asymptotic expansion for the integral in (\ref{psi}) is given by the contributions of its two real stationary phase points $\theta_\pm=\pm\,{\rm arccos}(\nu/x)$,
    \begin{equation}
    I(x,\nu)=\sqrt{\frac2\pi}
    \frac{\sin\left(\sqrt{x^2-\nu^2}-\nu\,{\rm arccos}\,\frac\nu{x}-\frac\pi4\right)}
    {(x^2-\nu^2)^{1/4}}+O(\hbar).   \label{sine}
    \end{equation}
This coincides with the asymptotic approximation of the Bessel function $J_{\nu}(x)$ of simultaneously large argument and order (``approximation by tangents" Eq. 8.453.1 of \cite{Gradstein}), so that in this domain of parameters the integral (\ref{I}) is in fact the ``one-loop" approximation of the Bessel function
    \begin{equation}
    I(x,\nu)=J_{\nu}(x)+O\left(\frac1x,\frac1\nu\right).
    \end{equation}
The phase of sine in (\ref{sine}) is of course the function $S(x,\nu)$ satisfying in terms of $\alpha$ and $p_{\phi}$ the Hamilton-Jacobi equation for (\ref{WdW2-phan})
    \begin{equation}
    -\frac19\,\left(\frac{\partial S}{\partial\alpha}\right)^2
    -p_{\phi}^2  +2V_0\,e^{6\alpha}=0.
    \end{equation}

However, in the underbarrier regime of the semiclassical approximation
    \begin{equation}
    x<\nu,           \label{underbarrier}
    \end{equation}
the integral is given asymptotically by contributions of the boundary points $\theta=\pm\pi/2$, because there are no stationary phase points between them. These contributions are $O(1/x,1/\nu)=O(\hbar)$ and go beyond the one-loop approximation. This means that the expression (\ref{psi}) does not reproduce correct semiclassical limit because it lacks in the underbarrier regime leading classical and one-loop terms.

On the other hand, underbarrier phenomena are usually described by the transition into a complex plane of time variable. This serves as a strong motivation to extend the range of the variable $\theta$ beyond $\pm\pi/2$ to the upper half of the complex plane,
    \begin{equation}
    \theta=\pm\frac\pi2+i\rho,\quad 0\leq\rho<\infty.
    \end{equation}
This, in its turn, corresponds to the extension of the range of the physical time variable from the segment $[-\sqrt{V_0},\sqrt{V_0}]$ to the full real axis
    \begin{equation}
    -\infty<T<\infty.
    \end{equation}
On the new regions with $|T|>\sqrt{V_0}$ the physical Hamiltonian (\ref{Ham-ph}) is imaginary, and the physical wavefunction becomes exponentially decaying  for $\rho\to\infty$,
    \begin{equation}
    \varPsi_{\rm phys}\big(\pm\sqrt{V_0}\cosh\rho,p_{\phi}\big) = \psi_0(p_{\phi})\,
    e^{\pm i\,\frac{\pi|p_{\phi}|}2-|p_\phi|\rho},
    \end{equation}
at the classically forbidden intervals of the whole time range $t=\sqrt{V_0}\sin(\pm\pi/2+i\rho)=\pm\sqrt{V_0}\cosh\rho$.

Thus, if we want to include in the Wheeler-DeWitt formalism description of underbarrier phenomena the generalized Fourier transform from $\varPsi_{\rm phys}$ to $\varPsi(\alpha,p_{\phi})$ defined by (\ref{psi})-(\ref{I}) should involve integration over the full real axis of $T$, or in terms of $\theta$
    \begin{eqnarray}
    &&\varPsi(\alpha,p_{\phi})
    =\sqrt{6\pi}\,
    \psi_{0}(p_{\phi})\,J(x,\nu),     \label{psiJ} \\
    &&J(x,\nu)=\frac1{2\pi}
    \int\limits_{-\pi/2+i\infty}^{\pi/2+i\infty}
    d\theta\,e^{-ix\sin\theta +i\nu\theta}. \label{J}
    \end{eqnarray}
Here the integration contour  runs vertically down from $\theta = -\pi/2+i\infty$ to $\theta = -\pi/2$, then along the real axis to $\theta = \pi/2$ and eventually goes vertically up to $\theta = \pi/2 + i\infty$. In the underbarrier range (\ref{underbarrier}) this integral has a saddle point $\theta_+=i\ln(\nu/x+\sqrt{\nu^2/x^2-1})$ which contributes the leading one-loop term
    \begin{equation}
    J(x,\nu)=\frac{e^{\sqrt{\nu^2-x^2}}}
    {\sqrt{2\pi}\,(\nu^2-x^2)^{1/4}}
    \left(\frac{x}{\nu+\sqrt{\nu^2-x^2}}\right)^\nu+O(\hbar)
    \end{equation}
missing in $I(x,\nu)$. This term is again the asymptotic expression for the Bessel function of large argument and order with $x<\nu$.

Moreover, as shown in Appendix B, the integral (\ref{J}) is exactly the representation of the Bessel function $J_{\nu}(x)$ for real positive argument $x$ and order $\nu$
    \begin{equation}
    J(x,\nu)=J_{\nu}(x), \quad x>0,\quad \nu>0, \label{int-B4}
    \end{equation}
so that finally
    \begin{equation}
    \varPsi(\alpha,p_{\phi}) = \sqrt{6\pi}\,\psi_{0}(p_{\phi})\,J_{|p_{\phi}|}
    \big(\sqrt{2V_0}\,e^{3\alpha}\big)+O(\hbar),\label{WdW-ph5}
    \end{equation}
which is one of the branches of the general solution of the Wheeler-DeWitt equation (\ref{WdW-sol-phan}), selected also by Hermiticity requirement.

Extension of integration range from (\ref{I}) to (\ref{J}) still might seem contrived because it implies violation of principles of physical reduction. This reduction starts entirely in classical terms and remains consistent unless the Faddeev-Popov factor (\ref{FP-ph}) degenerates to zero and physical evolution violates unitarity, that is for the domain $|T|<\sqrt{V_0}$. However, this domain does not cover full minisuperspace of $T$ or $\alpha$, and the artificial boundary at $T=\pm\sqrt{V_0}$ would mean a nonzero boundary value of $\varPsi(\alpha,p_{\phi})$ at $\alpha\to -\infty$ (or $x=0)$, because the integral (\ref{I}) has a finite limit $I(0,\nu)=O(1)$ rather than exponential falloff $x^\nu\sim e^{3\alpha}$. Therefore, no Hermiticity properties of momentum operators in reduced superspace $-\sqrt{V_0}<T<\sqrt{V_0}$ can be spoken of.
\footnote{Not to mention that with the integration limits $T=\pm\sqrt{V_0}$ ``integration by parts" of $(\partial/i\partial T)^{-1/2}$ in the derivation of (\ref{XXXX}) (cf. discussion of Eq.(\ref{X})) is impossible
without uncontrollable boundary terms.}
This strongly suggests extension of minisuperspace of $T$ up to infinity. This extension embraces a classically forbidden domain by the price of adding non-unitary dynamics in the physical sector (complex time or imaginary physical Hamiltonian) but retaining real values of the minisuperspace variable $T$ and $\alpha$.

\section{Conclusions}

Dirac quantization of constrained systems does not form a closed physical theory. Quantum Dirac constraints, which in gravity theory context form the set of Wheeler-DeWitt equations, have many more solutions than those corresponding to the physical setting of the problem. The way to select physically meaningful solutions matching with quantum initial value data may consist in reduced phase space quantization. Performing reduction to a physical sector results after quantization in the physical wavefunction that can be raised to the level of the wavefunction in superspace of 3-metric and matter fields. This superspace wavefunction forms a subset of solutions of Wheeler-DeWitt equations parameterized by the physical initial data.

This program can explicitly be realized in spatially homogeneous (minisuperspace) setup with one Hamiltonian constraint, and this was demonstrated for three simple but essentially nonlinear models: flat FRW cosmology with a scalar field having a negative constant potential (or negative $\varLambda$-term), vanishing potential and, finally, a phantom scalar field with a positive constant potential. Quite remarkably, the resulting selection rules for solutions of the minisuperspace Wheeler-DeWitt equation leave us only with those of its wavefunctions which guarantee Hermiticity of canonical phase space operators of the theory. This property is an important ingredient of the Dirac quantization scheme, but it is not a priori guaranteed to be true -- generic set of solutions of the Wheeler-DeWitt equation is not square integrable and violates Hermiticity of canonical momenta operators.

Central point of physical reduction is the choice of temporal gauge condition, or the choice of time $T$ as a function of phase space variables, which allows one to disentangle physical degrees of freedom, their Hamiltonian and unitarily evolving quantum state -- the physical wavefunction. Consistency of this gauge fixing procedure, or nondegeneracy of the corresponding Faddeev-Popov operator, is equivalent to the requirement that monotonically growing time variable should be in one-to-one correspondence with evolving state of the system. The hint for construction of $T$ comes from considering two different types of the classical cosmological evolution - with or without turning  points, i.e. the points of maximal expansion or the points of minimal contraction (bounces).

In the case, when the classical evolution  represents only expansion or contraction, it is sufficient to use gauges which fix the so called intrinsic time parameter, i.e. a parameter, which depends on the cosmological scale factor and is independent of its conjugated momentum. Two different physical Hamiltonians, as solutions of the quadratic Hamiltonian constraint equation, and the corresponding two physical wavefunctions  arise in this case. The latter additively enter their Wheeler-DeWitt counterpart in minisuperspace and give its two independent branches without any selection rule. This, however, does not contradict Hermiticity requirement, because both branches turn out to be square integrable and admit integration of the derivatives by parts -- Hermiticity of canonical momenta. This happens in the scalar field model with a vanishing potential.

Qualitatively different situation arises in models whose evolution includes turning points -- for a constant negative potential and for a phantom field with a positive potential. In this case, the intrinsic time gauges are inadequate, because a single value of the cosmological scale factor labels two different states - those of expansion and contraction. Instead, one  should use  extrinsic time -- the function of the Hubble parameter whose values at least classically are in one-to-one correspondence with consecutive moments of the cosmological evolution, including the turning points and bounces. We make physical reduction with this Hubble time $T$, find the physical state evolving in this time and then raise it to the level of the wavefunction in minisuperspace of the scale factor $e^\alpha$ by a kind of a generalized Fourier transform from $T$-variable to $\alpha$-variable. Critical point of this procedure is a nontrivial selection rule -- the result selects one branch of t!
 he generic solution of the Wheeler-DeWitt equation, which is square integrable and satisfies Hermiticity requirement.

Note that the above method can also be extended to the degenerate case -- the minisuperspace FRW model without matter fields entirely driven by a cosmological constant, used in pioneering papers for the construction of the tunneling \cite{Vil,tun1} and no-boundary \cite{HH} wavefunctions of the Universe. Zero number of physical degrees of freedom in this case does not prevent us from applying the above procedure of raising the physical wavefunction to the minisuperspace level.
The resulting selection rule, however, turns out to be different from the outgoing waves version of the tunneling cosmological state \cite{Vil-bound}. Neither it is related to cosmological singularities or behavior of the wavefunction at infinity of the scale factor.  Rather, it is connected with the behavior in classically forbidden domains in minisuperspace -- the shadow regions behind the turning point. Namely, in such regions one of the branches of the general solution of the Wheeler-DeWitt equation is infinitely growing and, hence, should be discarded. In this respect our selection rule seems closer to the no-boundary proposal of Hartle and Hawking \cite{HH}, but a detailed comparison would require introduction of a spatial curvature and will be considered in the future publication.

As we saw, for a vanishing potential no selection rules are enforced. This is natural because in this case one does not have classically forbidden regions in minisuperspace. This sounds like classical-to-quantum correspondence in cosmology. The structure of a classical evolution  (presence or absence of turning points) determines the correct class of gauge conditions and absence or presence of selection rules for the solution of the Wheeler-DeWitt equation.\footnote{Another example of classical-to-quantum correspondence in cosmology was recently discussed in context of soft cosmological singularities \cite{Manti}.}

The formalism of quantum reduction to the physical sector \cite{BarvU,BKr,geom} is known only semiclassically (including tree-level exponential and one-loop prefactor). The obtained results hold also with the same precision. Moreover, conformity of reduced phase space quantization with the Dirac quantization was found in \cite{BarvU,BKr,geom} only in the classically allowed (overbarrier) semiclassical domain. It holds in the sense that unitary evolution in the physical sector was mapped onto semiclassical oscillating solutions of the Wheeler-DeWitt equation. The model of phantom scalar field shows that this mapping can be extended to the classically forbidden (underbarrier) domain with exponentially damped wavefunctions. This represents the extension beyond original principles of reduced phase space quantization, because in this domain it deals with non-unitary dynamics in the physical sector. In particular, this extension uses a complex physical Hamiltonian or complex physi!
 cal time and encounters degeneration of the Faddeev-Popov factor (\ref{FP-ph}) to zero at the boundary between the classically allowed domain and the forbidden one, $|T|=\sqrt{V_0}$. Nevertheless this extended quantum reduction provides important Hermiticity properties in Dirac quantization and selection of $L_2$-integrable Dirac-Wheeler-DeWitt wavefunctions. Moreover, as it was discussed in \cite{UFN} the caustic in the congruence of classical histories in superspace separating its overbarrier domains from the underbarrier ones always leads to vanishing Faddeev-Popov factor. This maintains the spirit of Euclidean quantum gravity -- the concept underlying the notion of the no-boundary wavefunction which describes both classically allowed and forbidden phases of the cosmological state by {\em real} superspace variables \cite{HH,Vil,tun1,tun-we}.

Our principal conclusion on conformity of physical reduction and Hermiticity selection rules was attained only in simple models. Consideration of more complicated cosmological models with a full set of (inhomogeneous) field-theoretical modes can pose additional problems. For example, the one-loop approximation raises the issue of correspondence between covariant calculations and those based on explicit reduction to physical degrees of freedom. This was intensively discussed in the cosmological context \cite{one-loop} and in the context of the vacuum energy calculation on the background of wormholes and gravastars \cite{Garattini}. This means that beyond minisuperspace approximation one should be more cautious with regard to physical reduction. However, semiclassical nature of the method which captures the effect of superspace boundaries and underbarrier domains gives a hope that our main  conclusion might be a generic feature of Dirac quantization, and we hope to study this !
 conjecture in the future.

\section*{Acknowledgments}
A.B. is grateful to the Section of INFN of Bologna and to the Department of Physics and Astronomy of the University of Bologna for hospitality during his visits to Bologna
in March of 2012 and in May of 2013. A.K. is grateful to A. Vilenkin for fruitful discussions and for hospitality during his visit to the Tufts Institute of Cosmology in September of 2013. The work of A.B. was partially supported by the RFBR grant No 14-01-00489. The work of A.K. was partially supported by the RFBR grant No 14-02-01173.

\appendix
\renewcommand{\thesection}{\Alph{section}}
\renewcommand{\theequation}{\Alph{section}.\arabic{equation}}

\section{Unitary canonical transformations}
Here we remind the formalism of the unitary transformation (\ref{Fourier00}) corresponding to the classical canonical transformation (\ref{cantran0})-(\ref{cantran5}). The latter implies that old and new canonical variables are related by
    \begin{equation}
    p_m=\frac{\partial F(q,\tilde q)}{\partial q^m},\quad
    \tilde p_m=
    -\frac{\partial F(q,\tilde q)}
    {\partial\tilde q^m}.
    \end{equation}
To guarantee Hermiticity of canonical operators at the quantum level this transformation should be specified by Weyl ordering in the right hand sides of these relations
    \begin{equation}
    \hat p_m=N_W\,\frac{\partial F(\hat q,\hat{\tilde q})}{\partial \hat q^m},\quad
    \hat{\tilde p}_m=
    -N_W\frac{\partial F(\hat q,\hat{\tilde q})}
    {\partial\hat{\tilde q}^m}.
    \end{equation}

The kernel of unitary transformation $\langle\,q\,|\,\tilde q\,\rangle$ can be found in the first (one-loop) order of $\hbar$-expansion by the following sequence of relations. First we rewrite Weyl normal ordering in terms of $q\tilde q$-ordering, when all operators $\hat q$ stand to the left of the operators $\hat{\tilde q}$
    \begin{eqnarray}
    &&\!\!\!\!\!\!N_W\,\frac{\partial F(\hat q,\hat{\tilde q})}{\partial \hat q^m}=
    N_{q\tilde q}\left(\frac{\partial F(\hat q,\hat{\tilde q})}{\partial \hat q^m}\right.\nonumber\\
    &&\!\!\!\!\!\!\quad\quad\left.-\frac12\,[\,\hat q^n,\hat{\tilde q}^k\,]
    \frac{\partial^2}{\partial\hat q^n \partial\hat{\tilde q}^k}\, \frac{\partial F(\hat q,\hat{\tilde q})}{\partial \hat q^m}\right)+O(\hbar^2),
    \end{eqnarray}
where the commutator $[\,\hat q^n,\hat{\tilde q}^k\,]$ with the same precision is given by the Poisson bracket
    \begin{eqnarray}
    &&[\,\hat q^n,\hat{\tilde q}^k\,]=i\hbar\,\{\,\hat q^n,\hat{\tilde q}^k\}+O(\hbar^2)=i\hbar\,\frac{\partial\hat{\tilde q}^k}{\partial\hat p_n}+O(\hbar^2)\nonumber\\
    &&\qquad\qquad=
    i\hbar\,\left(\frac{\partial^2F(\hat q,\hat{\tilde q})}{\partial \hat q^n\partial\hat{\tilde q}^k}\right)^{-1}+O(\hbar^2).
    \end{eqnarray}
Therefore, since $\langle\,q\,|\,N_{q\tilde q}f(\hat q,\hat{\tilde q})\,|\,\tilde q\,\rangle=f(q,\tilde q)\,
\langle\,q\,|\,\tilde q\,\rangle$ for any function $f(\hat q,\hat{\tilde q})$ of noncommuting operators, we have
    \begin{eqnarray}
    \frac\hbar{i}\,\frac\partial{\partial q^m}\,
    \langle\,q\,|\,\tilde q\,\rangle&=&
    \langle\,q\,|\,\hat p_m\,|\,\tilde q\,\rangle
    \nonumber\\
    &=&\frac{\partial}{\partial q^m}
    \left(F
    -\frac{i\hbar}2\,\ln\,{\rm det}
    \frac{\partial^2 F}{\partial q^n \partial\tilde q^k}\right)\langle\,q\,|\,\tilde q\,\rangle\nonumber\\
    &&\quad+O(\hbar^2),
    \end{eqnarray}
whence 
    \begin{eqnarray}
    \!\!\!\langle\,q\,|\,\tilde q\,\rangle=
    \left|\,{\rm det}\,\frac1{2\pi\hbar}
        \frac{\partial^2F(q,\tilde q)}{\partial q^i\;\partial \tilde q^k}\,\right|^{1/2}
        e^{\frac{i}{\hbar}\,F(q,\tilde q)}
        +O(\hbar),                           \label{A}
        \end{eqnarray}
where the normalization coefficient follows from unitarity requirement, $\int d\tilde q\,\langle\,q\,|\,\tilde q\,\rangle \langle\,\tilde q\,|\,q'\,\rangle=\delta(q-q')$. In fact, this expression represents the Pauli-Van Vleck-Morette equation for the kernel of unitary evolution from $\tilde q$ to $q$ when $F(q,\tilde q)$ is identified with the relevant Hamilton-Jacobi function. 

The case of contact canonical transformations (\ref{Fourier01}), $q\to\tilde q=\tilde q(q)$, requires somewhat different consideration because its classical generating function relates old coordinates $q$ with new momenta $\tilde p$,
    \begin{equation}
    dF(q,\tilde p)=p\,dq+\tilde q\,d\tilde p,
    \end{equation}
and equals $F(q,\tilde p)=\tilde p_m\tilde q^m(q)$. Therefore, the kernel of transformation from $\hat q$ to $\hat{\tilde p}$ -- the generalized Fourier transform -- reads according to (\ref{A})
    \begin{eqnarray}
    \!\!\!\langle\,q\,|\,\tilde p\,\rangle=
    \left|\,{\rm det}\,\frac1{2\pi\hbar}
        \frac{\partial\tilde q}{\partial q}\;\right|^{1/2}
        e^{\frac{i}{\hbar}\,\tilde p\,\tilde q(q)}
        +O(\hbar),                           \label{A1}
        \end{eqnarray}
and the coordinate representation kernel
    \begin{eqnarray}
    \!\!\!\langle\,q\,|\,\tilde q\,\rangle&=&\frac1{\sqrt{2\pi\hbar}}\int d\tilde p\,
    \langle\,q\,|\,\tilde p\,\rangle\,
    e^{-\frac{i}{\hbar}\,\tilde p\,\tilde q}\nonumber\\
    &=&\left|\,{\rm det}\,
        \frac{\partial\tilde q}{\partial q}\;\right|^{1/2}
        \delta(\tilde q(q)-\tilde q),
        \end{eqnarray}
yields the transformation law (\ref{Fourier01}). 

\section{Integral representations for Bessel and modified Bessel functions}
Eq. (\ref{Watson2}) is based on the integral representation for the MacDonald function $K_{\nu}(x)$ \cite{Watson}:
    \begin{equation}
    K_{\nu}(x) = \frac12e^{-\frac12 \nu \pi i} \int_{-\infty}^{\infty} dt\,
    e^{-ix\sinh t -\nu t},                \label{Mac0}
    \end{equation}
which is valid for real positive argument $x$ and order $\nu$ belonging to the range $-1 < {\rm Re}\,\nu < 1$. Using the symmetry $K_{-\nu}(z) = K_{\nu}(z)$ and changing the sign of the integration parameter $t$ we immediately come to
    \begin{equation}
    K_{\nu}(x)=\frac12 e^{\frac12 \nu\pi i}
    \int_{-\infty}^{\infty}dt\,e^{ix\sinh t
    -\nu t},                            \label{Watson21}
    \end{equation}
which is just Eq. (\ref{Watson2}).

To derive the exact integral representation (\ref{J}) for the Bessel function (\ref{int-B4}) we start with Eq.
8.412.6 of \cite{Gradstein}
    \begin{equation}
    J_{\nu}(z) = \frac{1}{2\pi}\int\limits_{-\pi +i\infty}^{\pi+i\infty}d\theta\,
    e^{-iz\sin\theta +i\nu\theta},
    \ {\rm Re}\  z > 0.         \label{Bes-con}
    \end{equation}
The contour of integration here starts at $\theta = -\pi+i\infty$, runs vertically down to $\theta = -\pi$, follows along the real axis to the point $\theta = \pi$ and then goes vertically up to $\theta = \pi + i\infty$.

The integral (\ref{J}) has the integration contour similar to that of (\ref{Bes-con}) with the points $\theta = \pm\pi$ replaced respectively by $\theta =\pm\pi/2$. Besides, the parameters $x$ and $\nu$ in the integrand of (\ref{J}) are both real and positive. We want to show that this integral also gives the Bessel function $J_{\nu}(x)$. Since
    \begin{eqnarray}
    &&\!\!\!\!\!\!\!\!\!\!\!\!\!\!\!\!\!\!\!\!
    \int\limits_{-\pi/2 +i\infty}^{\pi/2+i\infty}d\theta\,e^{-ix\sin\theta +i\nu\theta}=
    \left(\int\limits_{-\pi +i\infty}^{\pi+i\infty}\right.
    \nonumber\\
    &&\!\!\!\!\!\!\!\!\!\!\!\!\qquad\left.-\int\limits_{-\pi +i\infty}^{-\pi/2+i\infty}-\int\limits_{\pi/2 +i\infty}^{\pi+i\infty}\right)\,d\theta\,e^{-ix\sin\theta +i\nu\theta}
    \label{int-B1}
    \end{eqnarray}
this statement reduces to the fact that the last two integrals in the right-hand side of this equation vanish.  To prove it consider the first of these two integrals. It coincides with the integral over the horizontal segment of
    \begin{equation}
    \theta = \beta + i\Lambda,
    \label{interval-B}
    \end{equation}
with the real part of $\theta$ running between $-\pi$ and $-\pi/2$, $-\pi \leq \beta \leq -\pi/2$, and the constant imaginary part $\Lambda$ tending to infinity, $\Lambda \rightarrow \infty$. At this segment the exponential of the integrand
    \begin{equation}
    e^{-ix\sin\theta +i\nu\theta}=e^{-ix\sin\beta\cosh\Lambda
    +x\cos\beta\sinh\Lambda+i\nu\beta-\nu\Lambda}.
    \label{int-B2}
    \end{equation}
is dominated by a large real part $x\cos\beta\sinh\Lambda-\nu\Lambda$, which is negative in view of $\cos\beta \leq 0$ and $\nu>0$. Therefore, in the limit $\Lambda \rightarrow \infty$ the integrand uniformly tends to zero, and the integral $\int_{-\pi +i\infty}^{-\pi/2+i\infty}e^{-ix\sin\theta +i\nu\theta}d\theta$ vanishes. The same is true for the second integral
$\int_{\pi/2 +i\infty}^{\pi+i\infty}e^{-ix\sin\theta +i\nu\theta}d\theta$. Thus, for real positive $x$ and $\nu$ the integral (\ref{J}) coincides with the integral representation of the Bessel function (\ref{Bes-con}) for $z=x$.

\end{document}